# A systematic approach to improving the reliability and scale of evidence from health care data

Martijn J. Schuemie [1,2,*], Patrick B. Ryan [1,2,3], George Hripcsak [1,3,4], David Madigan [1,5], Marc A. Suchard [1,6,7,8]

1. Observational Health Data Sciences and Informatics (OHDSI), New York, NY 10032
2. Epidemiology Analytics, Janssen Research and Development, Titusville, NJ 08560
3. Department of Biomedical Informatics, Columbia University Medical Center, New York, NY 10032
4. Medical Informatics Services, New York-Presbyterian Hospital, New York, NY 10032
5. Department of Statistics, Columbia University, New York, NY 10027
6. Department of Biomathematics, University of California, Los Angeles, CA 90095
7. Department of Biostatistics, University of California, Los Angeles, CA 90095
8. Department of Human Genetics, University of California, Los Angeles, CA 90095

*To whom correspondence should be addressed. E-mail: schuemie@ohdsi.org

## Abstract

Concerns over reproducibility in science extend to research using existing healthcare data; many observational studies investigating the same topic produce conflicting results, even when using the same data. To address this problem, we propose a paradigm shift. The current paradigm centers on generating one estimate at a time using a unique study design with unknown reliability and publishing (or not) one estimate at a time. The new paradigm advocates for high-throughput observational studies using consistent and standardized methods, allowing evaluation, calibration, and unbiased dissemination to generate a more reliable and complete evidence base. We demonstrate this new paradigm by comparing all depression treatments for a set of outcomes, producing 17,718 hazard ratios, each using methodology on par with state-of-the-art studies. We furthermore include control hypotheses to evaluate and calibrate our evidence generation process. Results show good transitivity and consistency between databases, and agree with four out of the five findings from clinical trials. The distribution of effect size estimates reported in literature reveals an absence of small or null effects, with a sharp cutoff at p = 0.05. No such phenomena were observed in our results, suggesting more complete and more reliable evidence.

## Introduction

Great concern exists over reproducibility in science, with many scientists even using the term 'reproducibility crisis' (*1*). Low sample size, small effect sizes, data dredging (including P-hacking), conflicts of interest, large numbers of scientists working competitively in silos without combining their efforts, and so on, may conspire to dramatically increase the probability that a published finding is incorrect (*2*). Although many solutions have been proposed, including pre-registering studies, open science, team research, and better reporting, adoption of these solutions is still lacking (*3, 4*). Here we focus on reproducibility in observational research using existing health care data, where we believe a complementary solution is viable that would vastly improve reproducibility, while at the same time generate large amounts of reliable scientific evidence.

Existing health care data, such as claims and electronic health records, hold the promise of providing new insights to improve patient care. These data capture details of real-world experiences of patients and their encounters with the health care system, allowing the study of many types of therapies and revealing benefits received and harm done. Certainly, there exist limits to the range of questions that these data can answer as they are based on interactions with the healthcare system and depend on accurate recording of events. There is also an information asymmetry as 'harms' tend to come to medical attention and are easily reflected in these data while 'benefits' are often neither easily reflected in these data nor do they tend to drive patients to clinical encounters. Observational studies are more susceptible to bias, placing them lower in the hierarchy of clinical evidence than randomized clinical trials. Nonetheless, these data could yield a wealth of insights that go well beyond what can be explored through other sources of evidence.

Current observational research relies on one-off studies answering one question at a time with unique methodology and therefore unknown reliability, and disseminating these results (or not) one estimate at a time. Here we propose to unlock the potential of existing health care data by defining a high-throughput approach to observational research; we systematically compare all treatments for a given indication for a large set of outcomes captured in data from the Observational Health Data Science and Informatics (OHDSI) (*5*) research network. We adjust for measured confounders using propensity score stratification, a state-of-the-art confounder-adjustment strategy, but instead of the current practice of hand-picking covariates for the propensity model, we employ a completely data-driven approach to variable selection. In addition, uncertainty due to residual observational study bias, for example due to unmeasured confounders, is quantified by using control hypotheses (research questions with known answers). We employ both real negative control hypotheses (where the true hazard ratio is known to be 1) as well as synthetic positive control hypotheses (where the true hazard ratio is of known magnitude greater than 1), created by modifying negative controls. We subsequently express the observed uncertainty due to residual bias in calibrated confidence intervals (CIs) (*6*). We disseminate all results, thereby not only providing evidence at large scale, but also preventing publication bias. We demonstrate this new paradigm by comparing all treatments for depression for a large set of health outcomes using four large insurance claims databases, as depicted in Figure 1.

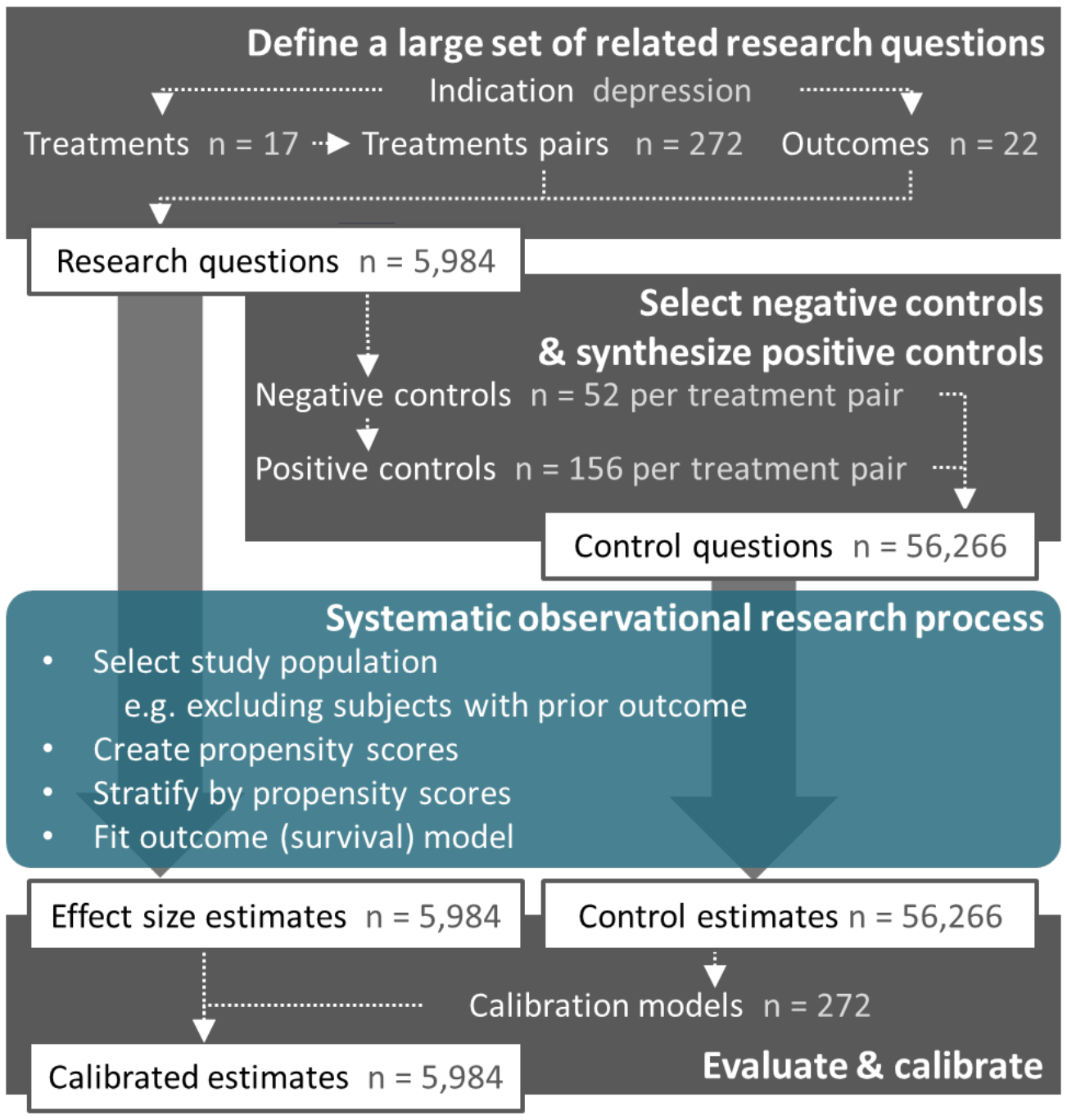

**Figure 1**. High-throughput observational study design with empirical calibration, applied to the comparison of depression treatments. We apply this design to four large insurance claims databases.

We evaluate our results in terms of transitivity and between-database consistency, and agreement with effects known from clinical trials. We also show that our distribution of estimates is markedly different from the distribution observed in literature, suggesting more complete and more reliable evidence.

## Results

### Example single research hypothesis

We demonstrate our high-throughput process by first showing the analysis for a single research question: the comparison of duloxetine to sertraline for the risk of stroke, using the Truven MarketScan Commercial Claims and Encounters (CCAE) database. We compare our approach to a previously published study by Lee et al.(*7*). Whereas that study compares new users of the entire drug classes to which these drugs belong, our analysis investigates new users of the two specific drugs. Both Lee et al. and our analysis require 12 months of continuous observation prior to treatment initiation, exclude people exposed to both drugs and people with prior strokes, and use stratification on the propensity score to address confounding. Follow-up is defined as starting on the day of treatment initiation and stopping on the day of the outcome, discontinuation of treatment (allowing a 30 day-gap between treatments), or disenrollment. Lee at al. hand-picked 74 covariates such as age, sex, and various selected drugs and diagnoses to create a propensity model. In contrast, we used a data-driven approach to generate a propensity model based on 59,038 covariates. Figure 2A shows our propensity score distribution across new users.

For many subjects, treatment assignment is highly dependent upon their baseline characteristics, indicating that the groups are fundamentally different and that without adjustment there is a high likelihood of confounding. On the other hand, Figure 2A also reveals substantial overlap, implying that propensity score adjustment should be able to make the groups equivalent, at least with regard to measured covariates. Indeed, Figure 2B shows that many covariates are imbalanced prior to adjustment, but after stratification all covariates have a standardized difference of mean smaller than 0.1, generally assumed to indicate adequate balance. This includes any covariates that experts might consider relevant such as comorbidities and current or prior medication use.

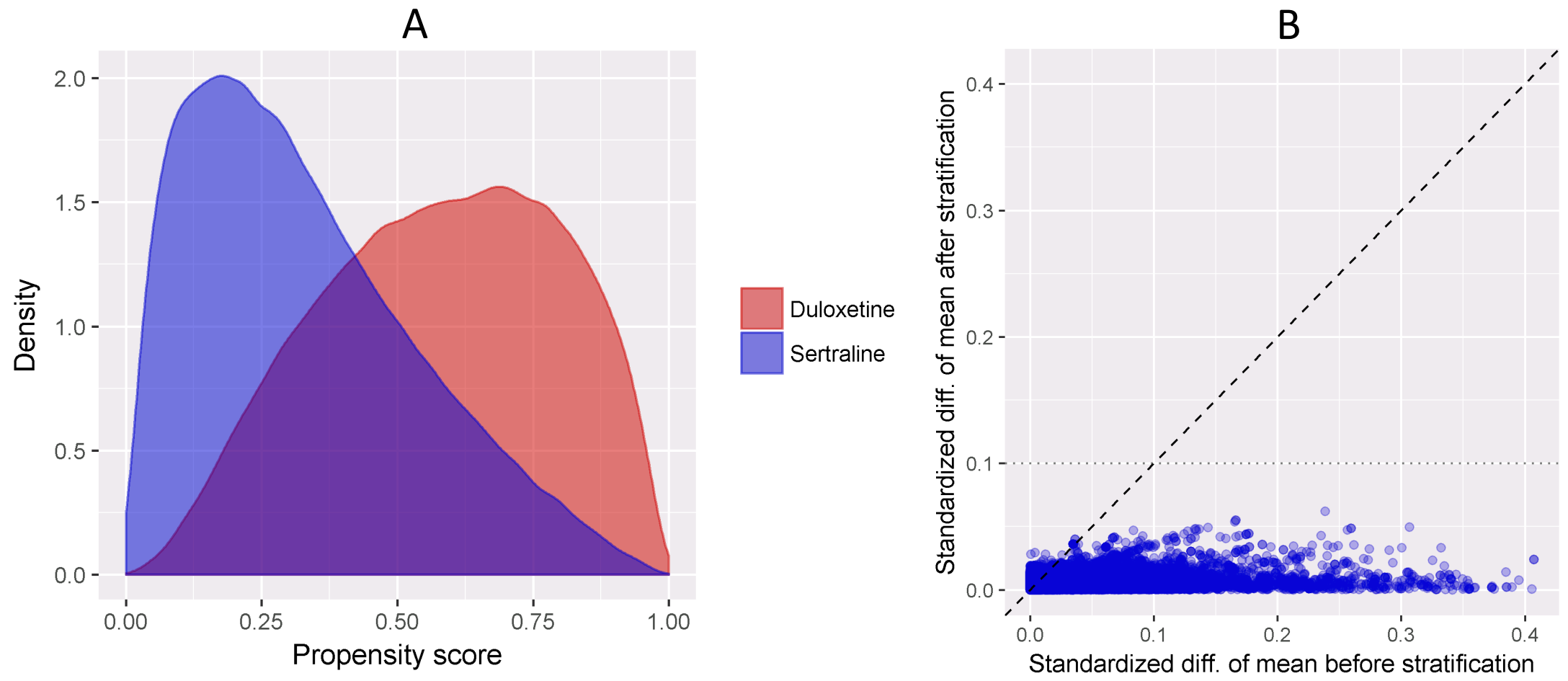

**Figure 2**. Cohort comparability and balance for duloxetine versus sertraline new users from the CCAE database. (A) Propensity score distributions for each cohort. (B) Absolute values of the standardized difference of the mean before and after stratification for the 59,038 covariates established at baseline.

For stroke risk, our analysis produces a propensity score-adjusted hazard ratio of 1.13 (95% CI: 0.81-1.61). This result stands in agreement with Lee et al.(*7*) who report an adjusted hazard ratio of 1.01 (95% CI: 0.90-1.12). Both studies also include sensitivity analyses that consider an alternative time-at-risk definition and show little variation in the estimate. We argue that the method used in both studies is of comparable rigor, and that our analysis meets the criteria for peer review, demonstrated by the publication of our studies using similar designs (*8-10*).

Figure 3 shows the estimates produced by applying the same analysis to a set of control outcomes (outcomes where the hazard ratio is known), while still comparing duloxetine to sertraline. This figure reveals the coverage of the uncalibrated 95% CI to be smaller than 95%. Calibrating the CIs using these observed operating characteristics restores near-nominal coverage.

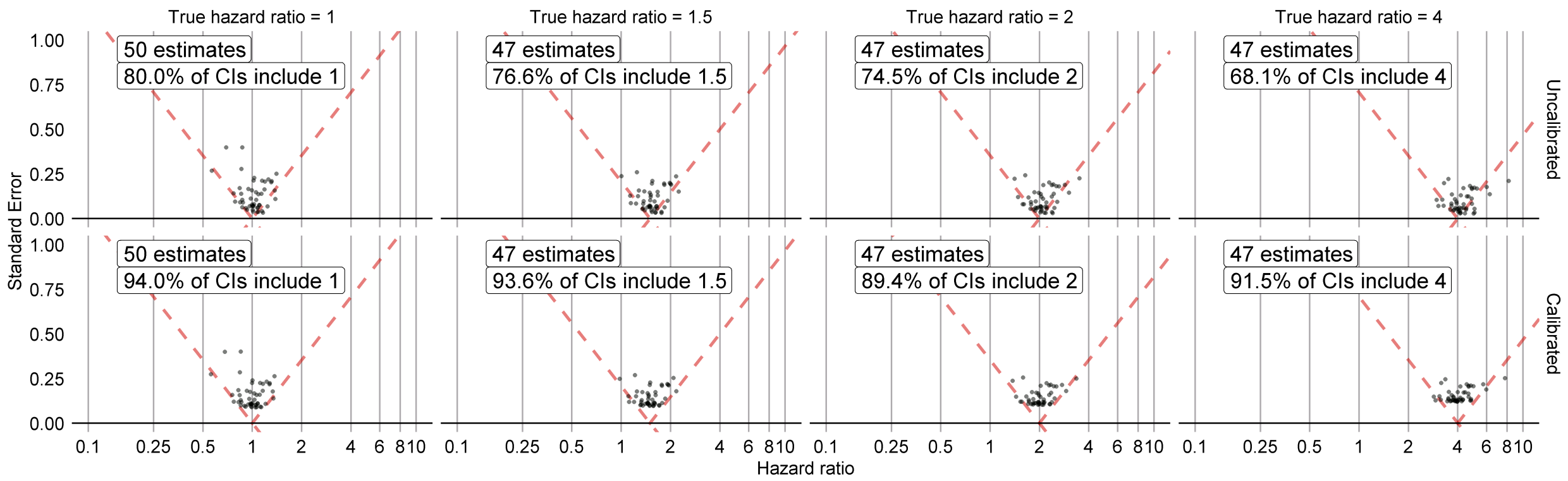

**Figure 3**. Evaluation of effect estimation between duloxetine and sertraline new users after stratification on the propensity scores before (top) and after (bottom) calibration. Each dot represents the hazard ratio and corresponding standard error for one of the negative (true hazard ratio = 1) or positive control (true hazard ratio > 1) outcomes.

Using the same calibration process, our hazard ratio for stroke becomes 1.11 (95% CI: 0.77-1.62), compared to the uncalibrated estimate of 1.13 (95% CI: 0.81-1.61). Although these estimates are similar, the empirical evaluation and calibration provide confidence that systematic error in our calibrated estimate remains small.

## Results of all comparisons

With four databases, the potential number of effect size estimates is (4 * 5,984 =) 23,936. We generate no risk estimate if at least one of the two treatment groups in a comparison contains less than 2,500 persons, so the final count is 17,718 estimates. The full set of results are available in the Supplementary Materials and can be explored online at http://data.ohdsi.org/SystematicEvidence. The results of our evaluation and calibration using control outcomes, verified by cross-validation can be found in the Supplementary Materials. The distribution of calibrated effect size estimates is also shown in figure 5C.

## Effect transitivity

If drug A has a statistically significant higher risk than drug B for a particular outcome, and drug B has a statistically significant higher risk than C for that same outcome, we expect A to have a statistically significant higher risk than C. In total, we identified 755 such A-B-C combinations, of which for 722 triplets (96%) the transitivity property held.

## Between-database consistency

Our previous work has suggested remarkably high heterogeneity when uncalibrated but identical observational study designs are implemented in different databases (*11*). In the present context, ideally, calibrated effects estimated across the four observational databases would be relatively consistent. For the 2,570 target-comparator-outcome triplets having sufficient data in all four databases, we compute the $I^2$ heterogeneity metric (*12*). An $I^2$ of zero means no between-database heterogeneity is observed. Across databases, 83% of calibrated estimates have an $I^2$ below 0.25; see Figure 4 for a complete histogram. In contrast, and in line with our previous work, only 58% of the estimates have an $I^2$ below 0.25 when no calibration is applied.

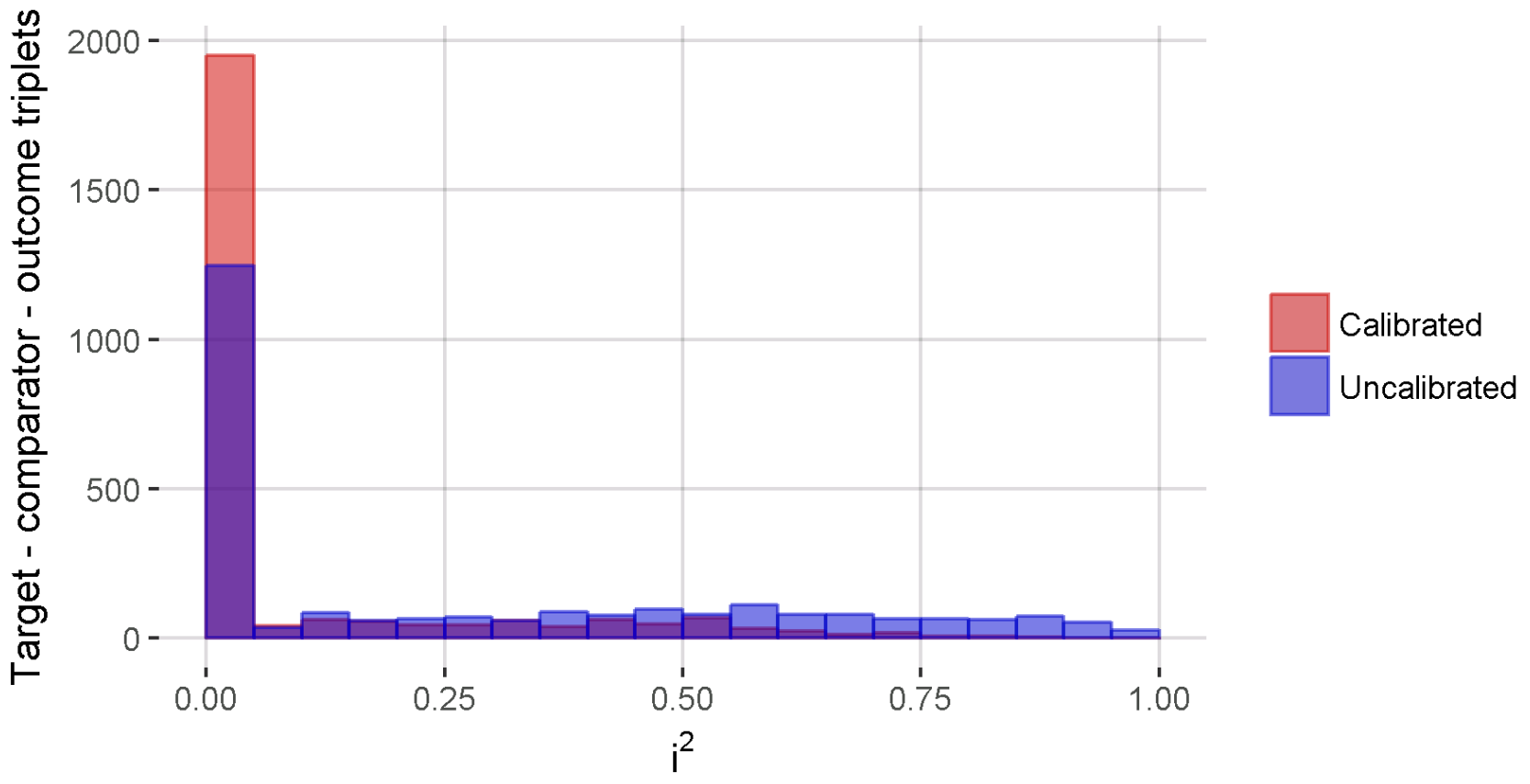

**Figure 4.** $I^2$ distribution for all 2,570 target-comparator-outcome triplets for which there was enough data in all four databases. Blue shows the distribution before calibration, red shows the distribution after calibration.

## Consistency with established knowledge

An additional test of validity compares our results with the current literature. Gartlehner et al. (*13*) systematically review comparative effects of antidepressant treatments based on randomized clinical trials (RCTs) and observational studies. Five findings emerge from the RCTs: 1) sertraline has higher risk of diarrhea than comparators; 2) venlafaxine has higher risk of nausea than selective serotonin reuptake inhibitors (SSRIs); 3) there is no difference in nausea between duloxetine and paroxetine or fluoxetine; 4) paroxetine has higher rate of sexual dysfunction than fluoxetine and sertraline; and 5) bupropion has lower incidence of sexual dysfunction than fluoxetine, paroxetine, and sertraline. Our result set correctly identified findings 1 through 4, as discussed in the Supplementary Materials—implying substantial but not perfect agreement with the literature. Supplementary Figure S3 also compares our results for finding 1 to estimates from RCTS reported in a systematic review on the topic (*14*), showing agreement as well as greater precision in our estimates due to the much larger sample size.

Finding 5 follows from five RCTs that demonstrated a significant lower rate of sexual adverse events in patients exposed to bupropion relative to SSRIs. Clinical guidelines suggest bupropion as an alternative treatment if a patient experiences sexual side effects with an SSRI medication (*15*) and other supporting trials recommend bupropion for patients for whom sexual dysfunction is a concern (*16*). From our result set, three databases return increased risks associated with bupropion relative to sertraline and fluoxetine. For example, in CCAE, the calibrated hazard ratio for decreased libido between bupropion and sertraline new users is 1.43 (1.09 – 1.89) and 1.42 (1.10 – 1.84) relative to fluoxetine new users. Channeling bias due to unmeasured baseline characteristics, such as sexual behavior, may explain this discordant finding.

## Comparing the distribution of estimates with the literature

To compare our approach to the current scientific process, we show the distribution of effect size estimates from observational studies reported in the literature (Figure 5A), the subset of estimates for depression treatments (Figure 5B), and compare it against estimates produced in the high-throughput study described in this paper (Figure 5C). At least three observations emerge from the current corpus of published observational studies. First, *the vast majority of effect estimates in literature abstracts (>80%) have a confidence interval (CI) that excludes one* (i.e., statistically significant effects at p<0.05). One explanation posits that researchers select hypotheses to test that have high *a priori* probabilities of being true. Another explanation is that observational studies are vulnerable to bias, for example due to confounding, selection bias, and measurement error, that can easily lead to statistically significant but erroneous results (*17*). Yet another explanation is that there is a tendency to only report results when the CI excludes one, resulting in publication bias. This ties into our second observation: *In evidence* *reported in the literature there is a sharp boundary between the regions where the CI does and does not include one*, suggesting that publication bias is pervasive. Third, when focusing on one specific area of interest (Figure 5B), in this case depression treatments, *the literature is sparse* compared to the more exhaustive approach taken in our study. Few of the questions that could have been asked are truly answered in literature, perhaps because the current process is too inefficient, slow, or because of the demonstrated publication bias.

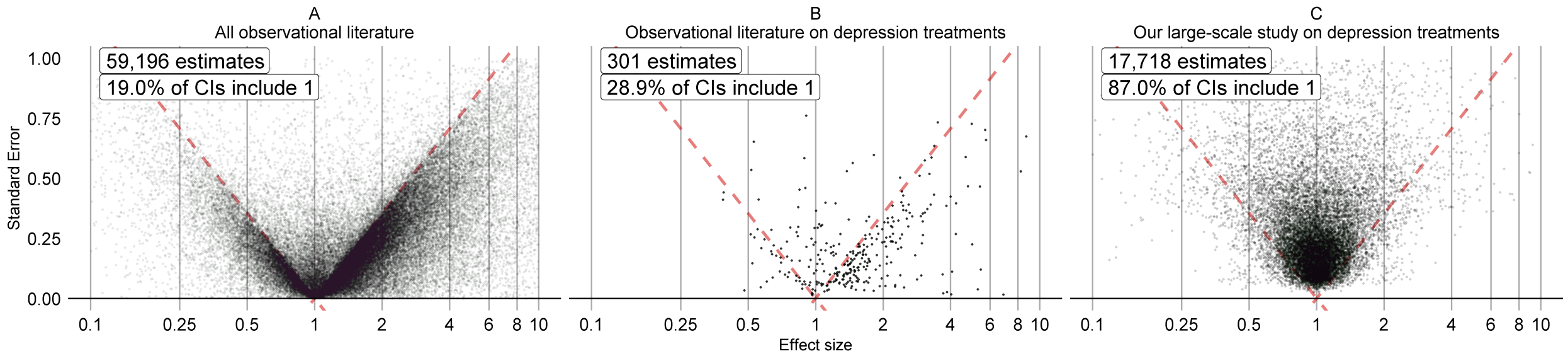

**Figure 5.** Effect size estimates from the literature (A, B) and the study described in this paper (C). Each dot represents a single estimate, such as relative risk, odds ratio, or hazard ratio, and corresponding standard error (linearly related to the width of the asymptotic confidence interval (CI)). Estimates below the red dashed line have a CI that excludes one, suggesting a non-null effect. Plot A shows estimates extracted from the abstracts of all observational research papers in MEDLINE, plot B shows only the subset of those that are related to depression treatments. Plot C shows estimated and calibrated hazard ratios for comparisons between depression treatments for various health outcomes of interest, generated from observational data in a single study using a systematic process. An online interactive visualization enables readers to explore these results in detail, including individual study artifacts for the estimates we generated (http://data.ohdsi.org/SystematicEvidence)

## Discussion

The distribution of estimates extracted from the literature (Figures 5A and 5B) exposes several concerns: Answers to many relevant questions are missing, either because they have not yet been investigated, or because publication bias hides effect sizes close to or equal to one. In addition, evidence that is present is unreliable for two reasons. One reason is the evident publication bias, making a high false-positive rate likely (*2*). In aggregate, published observational research is akin to data fishing at a massive proportion; by reporting primarily 'statistically significant' results and hiding others, spurious results due to random error appear legitimate because no adjustment is possible for the hidden multiple testing. The second reason is the proliferation of observational study bias (*17*). Indeed, observational research literature stands replete with multiple studies reporting statistically significant results in conflicting directions, even when employing the same data (*18-27*).

Applying a high-throughput observational study design can address these problems, as we have demonstrated in our example study comparing the effects of depression treatments: First, the evidence from such a study can be produced and disseminated as a single unit, and thereby prevent publication bias from reducing the validity of our results. Second, the inclusion of control hypotheses allows for evaluation of our study, measuring its operating characteristics such as coverage of the CI. We even use these measures to calibrate our results to restore nominal characteristics, as confirmed in our experimental results. Consequently, our estimates have a markedly different distribution compared with those found in the current literature, as demonstrated by comparing Figures 5A, 5B and 5C.

## Does quantity come at the cost of quality?

A potential criticism to our approach is that addressing many research questions at once is at odds with thoughtful study design for any single question, and therefore likely leads to lower quality research. However, each of our analyses is of high quality, sufficient to pass peer review as demonstrated by comparing our duloxetine-sertraline-stroke example to a published study and our prior publications using similar designs. Similarly, our evaluation using control hypotheses provides further confidence in the quality of our designs and goes above and beyond the recent tentative calls to include negative controls in observational studies (*28, 29*).

In fact, we believe unfettered freedom to customize a study for any research question is one of the main causes of the lack of reproducibility of observational study results, leading us to the situation portrayed in Figure 5A. Our challenge to the scientific community is to point out changes to our study design that researchers believe to be necessary when answering a particular question. Such changes should be evaluated objectively on their merit, for example using control hypotheses, and if proven to indeed improve quality, can be incorporated in a systematic way in the overall study design. Thus, science can move forward in a meaningful way, out of the current crisis.

## Limitations

We require our negative controls to be truly negative, but we rarely have definitive evidence of the absence of a causal relationship. We must assume that a lack of evidence of an effect for well-studied treatments and outcomes implies evidence of a lack of effect. In reality, some of our negative controls could prove to be positive at a future point in time.

In our evaluation and calibration procedure we require that the controls and the hypotheses of interest are exchangeable in certain aspects. We address this by choosing controls with the same target and comparator definitions, only differing in the outcome. However, negative controls could exhibit different bias than the outcomes of interest. Note that we do not assume the biases for these controls exactly equal the biases for the outcomes of interest, rather we assume only that biases draw from the same distribution. Unfortunately, we do not know for certain that this broad assumption holds. Furthermore, our positive controls fail to reflect bias due to unmeasured confounding other than that present for the negative controls on which they were based. However, we argue that detection of bias that may not fully represent all bias is better than ignoring bias completely.

A further limitation of observational research in general is that evidence can be generated only for those treatments and outcomes that are captured during interactions with the health care system and are reflected in the data. Some outcomes cannot be studied using these data. For example, we could not study reduction in depression symptoms as a possible outcome of treatment. Unmeasured confounding factors, as may have biased the estimate of the effect of bupropion on sexual dysfunction, remain a potential threat to the reliability of observational studies.

## How to use our results

Despite the limitations of observational data, they represent a critical component in improving the health care evidence base. Even though depression treatments have been extensively studied with hundreds of clinical trials, there is still much we do not know about the comparative effectiveness (including safety and tolerability) of alternative treatments. Evidence from our observational study can provide a reference to compare what we have learned in trials with what is observed in the real world. The evidence can also be a primary source when trials are unavailable, underpowered, or non-generalizable to the target population of interest.

We believe that our results should be used similarly to how one would use results currently scattered across observational research papers in the literature, which is typically a hypothesis-driven process. We purposely do ***not*** correct for multiple hypotheses in our results because that can only be done once a hypothesis or set of hypotheses is chosen. As when using results from literature, it is important to consider false positives when faced with multiple testing, and our results readily allow for adjustment for multiple testing, because we have disseminated all results. Note that such adjustments are not possible when using evidence scattered in literature, because many studies that should have been considered were never published due to publication bias. If readers dredge our results set looking for the most statistically significant ones, appropriate interpretation will require a multiplicity correction (e.g., Bonferroni or false discovery rate analysis). We believe that the value of our results, however, lies not in finding a few very significant ones, but in having available results that are poised to answer specific questions with as little bias as currently possible in observational research.

## Significance

To illustrate an example from our results, consider the outcome of suicidality. All antidepressants have FDA product labels that contain a black boxed warning for suicidal thinking and behavior. However, despite the considerable attention and public health importance of the outcome, "evidence from existing studies is insufficient to draw conclusions about the comparative risk of suicidality" (*13*). Our results readily provide evidence on this question, and to demonstrate this we consider SSRIs, the most prevalent class of antidepressants, and compare them to amitriptyline and bupropion, two other highly prevalent drugs that are not SSRIs. Our results in Figure 6 suggest that SSRIs (sertraline, fluoxetine, citalopram, escitalopram) may have an increased risk of suicidality relative to amitriptyline and bupropion. Evidence for other hypotheses similarly left unanswered by clinical trials could serve a valuable role to inform medical decision-making when weighing the collective benefits and harms of alternative treatments.

| Database | Target | Comparator | HR (95% CI) |
|---|---|---|---|
| CCAE | Citalopram | Amitriptyline | 2.60 (1.29-5.38) |
| MDCD | Citalopram | Amitriptyline | 3.91 (1.86-8.21) |
| MDCR | Citalopram | Amitriptyline | 2.39 (0.85-7.04) |
| Optum | Citalopram | Amitriptyline | 1.21 (0.81-1.83) |
| CCAE | Citalopram | Bupropion | 1.20 (1.05-1.38) |
| MDCD | Citalopram | Bupropion | 1.11 (0.84-1.46) |
| MDCR | Citalopram | Bupropion | 1.08 (0.70-1.66) |
| Optum | Citalopram | Bupropion | 1.19 (0.94-1.52) |
| CCAE | Escitalopram | Amitriptyline | 2.84 (1.45-5.52) |
| MDCD | Escitalopram | Amitriptyline | 2.17 (1.42-3.32) |
| MDCR | Escitalopram | Amitriptyline | 1.85 (0.74-4.62) |
| Optum | Escitalopram | Amitriptyline | 1.20 (0.85-1.75) |
| CCAE | Escitalopram | Bupropion | 1.28 (1.11-1.48) |
| MDCD | Escitalopram | Bupropion | 1.25 (0.89-1.71) |
| MDCR | Escitalopram | Bupropion | 1.02 (0.65-1.65) |
| Optum | Escitalopram | Bupropion | 1.27 (1.07-1.52) |
| CCAE | Fluoxetine | Amitriptyline | 1.77 (1.22-2.64) |
| MDCD | Fluoxetine | Amitriptyline | 2.07 (1.34-3.19) |
| MDCR | Fluoxetine | Amitriptyline | 1.98 (0.76-5.86) |
| Optum | Fluoxetine | Amitriptyline | 1.77 (1.25-2.56) |
| CCAE | Fluoxetine | Bupropion | 1.53 (1.27-1.85) |
| MDCD | Fluoxetine | Bupropion | 1.26 (1.03-1.55) |
| MDCR | Fluoxetine | Bupropion | 1.04 (0.66-1.67) |
| Optum | Fluoxetine | Bupropion | 1.58 (1.36-1.86) |
| CCAE | Sertraline | Amitriptyline | 3.57 (1.63-8.02) |
| MDCD | Sertraline | Amitriptyline | 1.79 (1.24-2.61) |
| MDCR | Sertraline | Amitriptyline | 2.16 (0.93-5.36) |
| Optum | Sertraline | Amitriptyline | 1.30 (0.94-1.84) |
| CCAE | Sertraline | Bupropion | 1.17 (0.94-1.48) |
| MDCD | Sertraline | Bupropion | 1.21 (0.92-1.59) |
| MDCR | Sertraline | Bupropion | 1.69 (1.17-2.43) |
| Optum | Sertraline | Bupropion | 1.22 (1.00-1.50) |

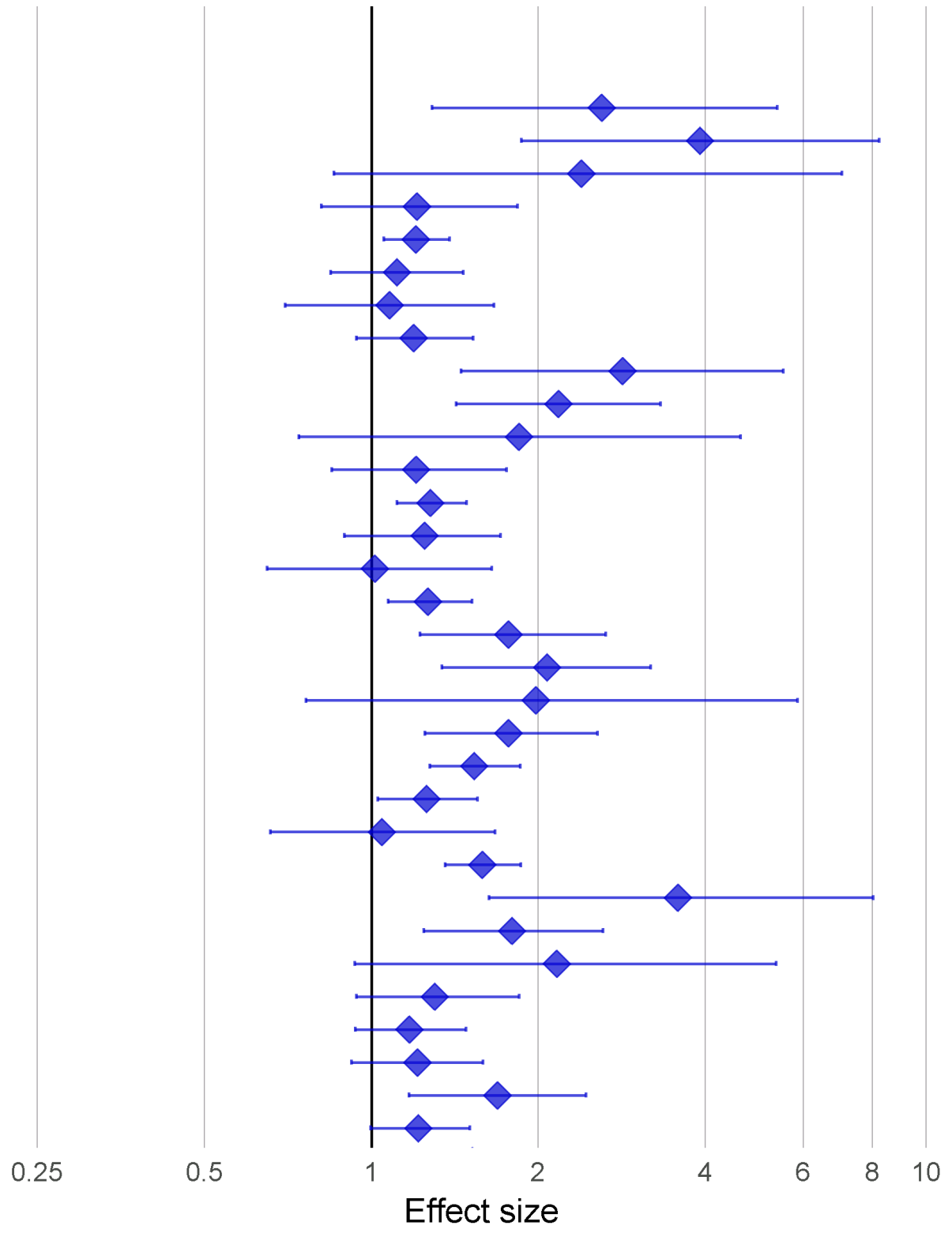

**Figure 6.** Subset of hazard ratio (HR) estimates and calibrated 95% confidence intervals (CI) generated in our study, for the outcome of suicide or suicidal ideation.

## Improving our study

While our choice of methods to address confounding (data-driven propensity score matching) and residual bias (confidence interval calibration) can be considered the current state-of-the-art, future studies could replace these with improved methods. In fact, we sincerely hope that observational researchers will move their focus from performing one-off studies to refining the high-throughput approach as described in this paper. Rather than each researcher working in isolation, we hope the scientific community will come together to build the process that generates evidence. To facilitate this, we have made all software necessary to execute this study available as open source.

## Conclusion

We propose a paradigm shift in how researchers generate and disseminate evidence from observational data. The current paradigm centers on generating one estimate at a time using a unique study design with unknown operating characteristics and publishing estimates one at a time through a dissemination process with clear limitations. The new paradigm advocates for larger scale studies that produce concurrent results for multiple hypotheses using consistent and standardized methods, allowing evaluation, calibration, and unbiased dissemination to generate a more reliable and complete evidence base than was previously possible. The results are poised for answering specific questions, able to be adjusted for multiple hypotheses as appropriate to the question at hand. Clinicians, regulators, and other medical decision makers can improve the care for patients by making well-informed decisions based on this evidence, and every treatment a patient receives becomes the basis for further evidence.

## Materials and Methods

### Comparison of depression treatments

As an example of our proposed high-throughput observational research we focus on the risk of specific outcomes across treatment choices for major depressive disorder. Depression is the leading cause of disability worldwide, affecting an estimated 350 million people globally (*30*), with multiple pharmacological and non-pharmacological treatments from which to choose. We identified 17 depression treatments to compare, and 22 outcomes of clinical interest (see Table 1). As a result, we have 17 * (17-1) * 22 = 5,984 research questions.

Our study follows a typical comparative effectiveness design (*31*), comparing a target treatment (T) to a comparator treatment (C) for the risk of an outcome (O). We create definitions of all Ts, Cs, and Os listed in Table 1, based on clinical knowledge and our understanding of the databases (see Supplementary Materials), and pre-specify the rules by which these definitions should be adapted for any specific combination of T, C, and O. For example, T and C are restricted to the calendar time when both treatments were recorded in the database, and people with prior O are removed from both T and C. Because of the observational nature of the study, subjects in T may differ from subjects in C in ways that could bias effect estimation. We apply a commonly used confounding adjustment strategy – stratification by propensity scores - to make the two cohorts more comparable. We define time-at-risk to start on the day of treatment initiation and stop when treatment stops, allowing for a 30-day gap in treatment continuation. We specify a sensitivity analysis where the time-at-risk is defined to stop at end of observation (end of enrollment or end of study period, whichever comes first). Hazard ratios are estimated using a Cox proportional model conditioned on the propensity score strata.

| Treatments of interest | Outcomes of interest |
|---|---|
| Amitriptyline | Acute liver injury |
| Bupropion | Acute myocardial infarction |
| Citalopram | Alopecia |
| Desvenlafaxine | Constipation |
| Duloxetine | Decreased libido |
| Electroconvulsive therapy | Delirium |
| Escitalopram | Diarrhea |
| Fluoxetine | Fracture |
| Mirtazapine | Gastrointestinal hemorrhage |
| Paroxetine | Hyperprolactinemia |
| Psychotherapy | Hyponatremia |
| Sertraline | Hypotension |
| Trazodone | Hypothyroidism |
| Venlafaxine | Insomnia |
| Vilazodone | Nausea |
| | Open-angle glaucoma |
| | Seizure |
| | Stroke |
| | Suicide & suicidal ideation |
| | Tinnitus |
| | Vent. arr. & sudden cardiac death |
| | Vertigo |

**Table 1.** Treatments and outcomes of interest

## Propensity score stratification

Adjustment for baseline confounders is done by fitting a propensity model and creating propensity scores (PS) for every pair of exposures. The propensity score is the probability of a subject receiving one treatment instead of the other, conditional on baseline characteristics (*32*). We create a data-driven process that entertains a large set of predefined baseline covariates—often tens of thousands—consistently for all combinations of T, C and O, and use the data to decide which combination of these characteristics are most predictive of the treatment assignment.

The following variables are included in all PS model:

- Demographics (age in 5-year increments, gender, race, ethnicity, year of index date, month of index date)
- Condition occurrence (one or more variables per diagnose code)
- Condition era (one or more variables per diagnose code)
- Condition group (one or more variables per MedDRA group or SNOMED groups)
- Drug exposure (one or more variables per drug code)
- Drug era (one or more variables per RxNorm ingredient)
- Drug group (one or more variables per ATC group)
- Procedure occurrence (one or more variables per procedure code)
- Observations (one or more variables per observation concept ID)
- Measurements (one or more variables per measurement concept ID, including variables for within / above / below normal range)
- Risk scores (including Charlson, DCSI, CHADS2, CHADS2VASc)

Variables with less than 100 non-zero values are discarded. For full details on the covariates used in our models please refer to FeatureExtraction package (https://github.com/OHDSI/FeatureExtraction)

The PS models are fitted using $L_1$ regularized regression (*33*), using 10-fold cross-validation to select the regularization parameter. These PS are used to stratify the target and comparator cohorts in 10 strata, and the proportional hazards outcome models are conditioned on the strata (*32*).

## Control hypotheses

We evaluate our process by applying it to research hypotheses where the truth is known with a high degree of certainty. Such a gold standard should include both negative and positive controls.

For comparative effectiveness studies, we define negative controls as TCO combinations where neither T nor C causes O and where therefore the true hazard ratio is equal to one. In practice, we identify negative controls by selecting exposures and outcomes that are well-studied, but for which no evidence in the literature or elsewhere suggests a relationship. For example, one negative control outcome is 'ingrown nail', because we firmly believe that no depression treatment causes ingrown nails. It is important to note that although there is no causal relationship, some antidepressants may be associated with ingrown nails, for example because the treatment is prescribed primarily for the elderly, where this condition is more prevalent. This allows us to test whether our confounding adjustment can correct for this confounding association, and produce estimates consistent with the null.

A candidate list of negative control outcomes was generated by identifying outcomes with no evidence of being causally related to any exposure of interest (*34*). This evidence was searched in literature through MeSH headings (*35*) and natural language processing (*36*), spontaneous reports of adverse events (*37*), and product labels in the US (*38*) and Europe (*39*). The candidate outcomes were then reverse sorted by prevalence in the observational databases and manually curated until a reasonably-sized set of negative controls was established. The final list of 52 negative control outcomes is provided in Table 2.

Positive controls in this case are outcomes believed to be caused by one exposure, but not the other. Unfortunately, real positive controls for observational research tend to be problematic for three reasons: First, when comparing the effect of two treatments there often is a paucity of positive controls relevant for that specific comparison. Second, even if positive controls are available, the magnitude of the effect size may not be known with great accuracy, and often depends on the population in which it is measured. Third, when treatments are widely known to cause a particular outcome, this will shape the behavior of physicians prescribing the treatment, for example by taking actions to mitigate the risk of unwanted outcomes, thereby rendering the positive controls useless as a means for evaluation (*40*). We therefore use synthetic positive controls (*6*), created by modifying a negative control through injection of additional, simulated occurrences of the outcome. To preserve (measured) confounding, simulated outcome occurrences are sampled from the probability distribution derived from a predictive model fitted on the data. These models use the same covariates as the propensity models as independent variables, and the occurrence of the negative control outcomes as the dependent variables. Target true hazard ratios for the positive control synthesis are 1.5, 2, and 4, so using the 52 negative controls we are able to construct 52 * 3 = 156 positive control outcomes for every comparison of two treatments. No negative control outcome model is fitted and no positive controls are created if there were less than 100 persons with the outcome across all exposures. No injection is performed if, for the exposure that is considered for injection, there were less than 25 persons with the outcome before injection.

| | |
|---|---|
| Acariasis | Ingrowing nail |
| Amyloidosis | Iridocyclitis |
| Ankylosing spondylitis | Irritable bowel syndrome |
| Aseptic necrosis of bone | Lesion of cervix |
| Astigmatism | Lyme disease |
| Bell's palsy | Malignant neoplasm of endocrine gland |
| Benign epithelial neoplasm of skin | Mononeuropathy |
| Chalazion | Onychomycosis |
| Chondromalacia | Osteochondropathy |
| Crohn's disease | Paraplegia |
| Croup | Polyp of intestine |
| Diabetic oculopathy | Presbyopia |
| Endocarditis | Pulmonary tuberculosis |
| Endometrial hyperplasia | Rectal mass |
| Enthesopathy | Sarcoidosis |
| Epicondylitis | Scar |
| Epstein-Barr virus disease | Seborrheic keratosis |
| Fracture of upper limb | Septic shock |
| Gallstone | Sjogren's syndrome |
| Genital herpes simplex | Tietze's disease |
| Hemangioma | Tonsillitis |
| Hodgkin's disease | Toxic goiter |
| Human papilloma virus infection | Ulcerative colitis |
| Hypoglycemic coma | Viral conjunctivitis |
| Hypopituitarism | Viral hepatitis |
| Impetigo | Visceroptosis |

**Table 2.** Negative control outcomes. Outcomes not believed to be caused by any of the exposures of interest.

With a gold standard in place, we evaluate whether our process produces results in line with the gold standard effect sizes. Importantly, we estimate CI coverage probability – the proportion of time that the CI contains the true value of interest. For example, we expect a 95% CI to cover the truth 95% of the time. We also apply a calibration procedure described elsewhere (*6*) that attempts to restore nominal coverage by adjusting the CIs, similarly to how one would calibrate a scale by using objects of known weight. In short, this procedure first estimates the distribution of systematic error using the observed estimates for negative and positive controls. We assume this distribution is Gaussian with a mean and log standard deviation linearly related to the true effect size. Using the estimated distribution, we then generate calibrated CIs considering both random and systematic error. Typically, but not necessarily, the calibrated CI is wider than the nominal CI, reflecting the problems unaccounted for in the standard procedure (such as unmeasured confounding, selection bias, and measurement error) but accounted for in the calibration.

## Observational databases

The analyses have been performed across a network of observational healthcare databases. All databases have been transformed into the OMOP Common Data Model, version 5. The complete specification for OMOP Common Data Model, version 5 is available at: https://github.com/OHDSI/CommonDataModel. The following databases have been included in this analysis:

- Truven MarketScan Commercial Claims and Encounters (CCAE)
- Truven MarketScan Medicare Supplemental Beneficiaries (MDCR)
- Truven MarketScan Multi-state Medicaid (MDCD)
- OptumInsight's de-identified Clinformatics™ Datamart (Optum)

### Truven MarketScan Commercial Claims and Encounters (CCAE)

CCAE is an administrative health claims database for active employees, early retirees, COBRA continues, and their dependents insured by employer-sponsored plans (individuals in plans or product lines with fee-for-service plans and fully capitated or partially capitated plans). As of 1 November 2016, CCAE contained 131 million patients with patient-level observations from January 2000 through July 2016. Source codes used in CCAE include: conditions- ICD-9-CM; drugs: NDC, HCPCS, ICD-9-CM; procedures: CPT-4, HCPCS, ICD-9-CM; lab: LOINC.

The ETL specification for transforming CCAE into the OMOP CDM is available at: https://github.com/OHDSI/ETL-CDMBuilder/tree/master/man/TRUVEN_CCAE_MDCR

**Truven MarketScan Medicare Supplemental Beneficiaries (MDCR)**

MDCR is an administrative health claims database for Medicare-eligible active and retired employees and their Medicare-eligible dependents from employer-sponsored supplemental plans (predominantly fee-for-service plans). Only plans where both the Medicare-paid amounts and the employer-paid amounts were available and evident on the claims were selected for this database. As of 1 November2016, MDCR contained 9.6 million patients with patient-level observations from January 2000 through July 2016. Source codes used in MDCR include: conditions- ICD-9-CM; drugs: NDC, HCPCS, ICD-9-CM; procedures: CPT-4, HCPCS, ICD-9-CM; lab: LOINC.

The ETL specification for transforming MDCR into the OMOP CDM is available at: https://github.com/OHDSI/ETL-CDMBuilder/tree/master/man/TRUVEN_CCAE_MDCR

**Truven MarketScan Multi-state Medicaid (MDCD)**

MDCD is an administrative health claims database for the pooled healthcare experience of Medicaid enrollees from multiple states. As of 1 November 2016, MDCD contained 21.6 million patients with patient-level observations from January 2006 through Dec ember 2014. Source codes used in MDCD include: conditions- ICD-9-CM; drugs: NDC, HCPCS, ICD-9-CM; procedures: CPT-4, HCPCS, ICD-9-CM; lab: LOINC.

The ETL specification for transforming MDCD into the OMOP CDM is available at: https://github.com/OHDSI/ETL-CDMBuilder/tree/master/man/TRUVEN_MDCD

**OptumInsight's de-identified Clinformatics™ Datamart (Optum)**

OptumInsight's de-identified Clinformatics™ Datamart (Eden Prairie,MN) is an administrative health claims database for members of United Healthcare, who enrolled in commercial plans (including ASO, 36.31M), Medicaid (prior to July 2010, 1.25M) and Legacy Medicare Choice (prior to January 2006, 0.36M) with both medical and prescription drug coverage. As of 1 November2016, Optum contained 74.7 million patients with patient-level observations from June 2000 through June 2016. Source codes used in Optum include: conditions- ICD-9-CM; drugs: NDC, HCPCS, ICD-9-CM; procedures: CPT-4, HCPCS, ICD-9-CM; lab: LOINC.

The ETL specification for transforming Optum into the OMOP CDM is available at: https://github.com/OHDSI/ETL-CDMBuilder/tree/master/man/OPTUM_EXTENDED

## Extraction from literature

Citations of observational studies were identified in PubMed using the following query:

```
("population-based" [Title/Abstract] OR observational [Title/Abstract] OR
pharmacoepidemiology [Title/Abstract]) AND (("Cohort Studies" [MeSH] OR
"cohort" [Title/Abstract] OR "propensity score" [Title/Abstract]) OR ("Case-
Control Studies" [MeSH] OR "case control" [Title/Abstract]) OR ("self
controlled case series" [Title/Abstract] OR ("sccs" [Title/Abstract] AND
"self-controlled" [Title/Abstract])) OR ("case-crossover" [Title/Abstract]) )
AND ("1900/01/01"[PDAT]:"3000/12/31"[PDAT])
```

In total, 102,874 citations were retrieved. The abstracts were automatically scanned for occurrences of the following regular expression:

```
"("+emPattern+" ?\\(|\\([^)]*"+emPattern+")[^(]*("+pValuePattern+"|"+ciPatter
n+")[^(]*\\)"
```

where

```
numberPattern = "[0-9][0-9]?[0-9]?\\.[0-9][0-9]?[0-9]?"
emPattern = "(odds ratio|o.r.|or|relative risk|r.r.|rr|hazard
ratio|h.r.|hr|hazard|rate ratio)([^0-9a-z]*| is | of )"+numberPattern
pValuePattern = "p ?[<=>] ?0?\\.[0-9][0-9]?[0-9]?"
ciPattern = numberPattern+" ?(-|to|,) ?" +numberPattern
```

In total, 59,196 estimates were found in 24,027 abstracts. The standard error was computed from either the confidence interval or p-value that was found in combination with an effect size estimate. If both a p-value and confidence interval were present, the confidence interval was used. The full list of estimates is provided in the supplementary materials (Data S2). To remove visual artifacts due to rounding, for visualization purposes only random noise was added to the estimates, confidence intervals, and p-values so that the noisy numbers would still round to the numbers reported in the abstracts. For example, a hazard ratio of 1.5 was converted to a random number between 1.450000001 and 1.549999999.

A subset of articles related to depression treat was identified using the PubMed query:

```
(depression OR antidepressant) AND ("serotonin reuptake inhibitors" OR
"tricyclic antidepressant" OR Bupropion OR Mirtazapine OR Trazodone OR
Desvenlafaxine OR duloxetine OR venlafaxine OR Citalopram OR Escitalopram OR
Fluoxetine OR Paroxetine OR Sertraline OR vilazodone OR Amitriptyline OR
Doxepin OR Nortriptyline or psychotherapy or "electroconvulsive therapy").
```

## Acknowledgments

This work was supported in part through National Science Foundation grants IIS 1251151 and DMS 1264153, and National Institutes of Health grants R01 LM06910 and U01 HG008680. We would like to thank Christian Reich, Jon Duke, and Paul Stang for their constructive comment on this paper.

# Supplementary materials

## Document S1. Exposure definitions

The target and comparator groups consisted of new users of the treatments listed in Table 1, which are identified using the codes specified in tables S1.1-S1.4. For both cohorts we restrict to people with a prior diagnosis of depression, and no prior history of bipolar disorder or schizophrenia.

| Concept ID | Concept Name |
|---|---|
| 4152280 | Major depressive disorder |
| 435783 | Schizophrenia |
| 436665 | Bipolar disorder |

**Table S1.1.** Concepts used to identify prior history of depression, schizophrenia, and bipolar disorders. All descendants of these concepts are also considered.

| Concept ID | RxNorm ID | Concept Name |
|---|---|---|
| 710062 | 704 | Amitriptyline |
| 750982 | 42347 | Bupropion |
| 797617 | 2556 | Citalopram |
| 717607 | 734064 | Desvenlafaxine |
| 738156 | 3638 | Doxepin |
| 715259 | 72625 | duloxetine |
| 715939 | 321988 | Escitalopram |
| 755695 | 4493 | Fluoxetine |
| 725131 | 15996 | Mirtazapine |
| 721724 | 7531 | Nortriptyline |
| 722031 | 32937 | Paroxetine |
| 739138 | 36437 | Sertraline |
| 703547 | 10737 | Trazodone |
| 743670 | 39786 | venlafaxine |
| 40234834 | 1086769 | vilazodone |

**Table S1.2.** The 15 included drugs and their concept identifiers. All descendants of these concepts are also considered.

| Concept ID | Concept Name |
|---|---|
| 4119335 | Analytical psychology |
| 4084202 | Anti-criminal psychotherapy |
| 4079608 | Anti-suicide psychotherapy |
| 4048385 | Brief group psychotherapy |
| 4295027 | Brief solution focused psychotherapy |
| 4299728 | Client-centered psychotherapy |
| 4164790 | Conjoint psychotherapy |
| 4208314 | Couple psychotherapy |
| 4083706 | Crisis intervention |
| 4083131 | Daily life psychotherapy |
| 4121662 | Developmental psychodynamic psychotherapy |
| 4226276 | Eclectic psychotherapy |
| 4258834 | Educational psychotherapy |
| 4148765 | Encounter group therapy |
| 2007747 | Exploratory verbal psychotherapy |
| 4137086 | Expressed emotion family therapy |
| 4048387 | Expressive psychotherapy |
| 4173581 | Extended family therapy |
| 46286403 | Family intervention for psychosis |
| 2213546 | Family psychotherapy (conjoint psychotherapy) (with patient present) |
| 4028920 | Family psychotherapy procedure |
| 46286330 | Focal psychodynamic therapy |
| 4226275 | Formal psychological therapy |
| 45765516 | Functional family therapy |
| 4079939 | Functional psychotherapy |
| 4079500 | General psychotherapy |
| 4117915 | Generic Jungian-based therapy |
| 4100341 | Group analytical psychotherapy |
| 44808677 | Group cognitive behavioural therapy |
| 4136352 | Group marathon therapy |
| 4268909 | Group primal therapy |
| 4296166 | Group psychotherapy |
| 2213548 | Group psychotherapy (other than of a multiple-family group) |
| 2617477 | Group psychotherapy other than of a multiple-family group, in a partial hospitalization setting, approximately 45 to 50 minutes |
| 4196062 | Group reassurance |
| 2213554 | Individual psychophysiological therapy incorporating biofeedback training by any modality (face-to-face with the patient), with psychotherapy (eg, insight oriented, behavior modifying or supportive psychotherapy); 30 minutes |
| 2213555 | Individual psychophysiological therapy incorporating biofeedback training by any modality (face-to-face with the patient), with psychotherapy (eg, insight oriented, behavior modifying or supportive psychotherapy); 45 minutes |
| 4088889 | Individual psychotherapy |
| 2007730 | Individual psychotherapy |
| 4103512 | Interactive group medical psychotherapy |
| 2617478 | Interactive group psychotherapy, in a partial hospitalization setting, approximately 45 to 50 minutes |
| 4221997 | Interactive individual medical psychotherapy |
| 40482841 | Interpersonal psychotherapy |
| 4119334 | Jungian-based therapy |
| 4118797 | Long-term exploratory psychotherapy |
| 4118798 | Long-term psychodynamic psychotherapy |
| 44792695 | Marital psychotherapy |
| 2213547 | Multiple-family group psychotherapy |
| 4118800 | Narrative family psychotherapy |
| 4242119 | Occupational social therapy |
| 2007749 | Other individual psychotherapy |
| 2007750 | Other psychotherapy and counselling |
| 45887728 | Other Psychotherapy Procedures |
| 45763911 | Parent-infant psychotherapy |
| 2007746 | Play psychotherapy |
| 4083133 | Potential suicide care |
| 4084195 | Provocative therapy |
| 2213544 | Psychoanalysis |
| 2007731 | Psychoanalysis |
| 4114491 | Psychoanalytic and psychodynamic therapy |
| 4202234 | Psychodrama |
| 2007763 | Psychodrama |
| 4199042 | Psychodynamic psychotherapy |
| 4128268 | Psychodynamic-interpersonal psychotherapy |
| 4118801 | Psychotherapeutic approaches using specific settings |
| 4327941 | Psychotherapy |
| 4083129 | Psychotherapy - behavioral |
| 4079938 | Psychotherapy - cognitive |
| 45889353 | Psychotherapy for crisis |
| 45888237 | Psychotherapy for Crisis Services and Procedures |
| 43527991 | Psychotherapy for crisis; each additional 30 minutes (List separately in addition to code for primary service) |
| 43527990 | Psychotherapy for crisis; first 60 minutes |
| 45887951 | Psychotherapy Services and Procedures |
| 2108571 | Psychotherapy services provided (MDD, MDD ADOL) |
| 43527986 | Psychotherapy, 30 minutes with patient and/or family member |
| 43527987 | Psychotherapy, 30 minutes with patient and/or family member when performed with an evaluation and management service (List separately in addition to the code for primary procedure) |
| 43527904 | Psychotherapy, 45 minutes with patient and/or family member |
| 43527988 | Psychotherapy, 45 minutes with patient and/or family member when performed with an evaluation and management service (List separately in addition to the code for primary procedure) |
| 43527905 | Psychotherapy, 60 minutes with patient and/or family member |
| 43527989 | Psychotherapy, 60 minutes with patient and/or family member when performed with an evaluation and management service (List separately in addition to the code for primary procedure) |
| 4148398 | Psychotherapy/sociotherapy |
| 4083130 | Rehabilitation for disabling psychiatric problem |
| 44791916 | Relationship psychosexual therapy |
| 4265313 | Relationship psychotherapy |
| 4084201 | Samaritans advisory service |
| 4233181 | Sensate focus technique |
| 4272803 | Sexual psychotherapy |
| 4035812 | Sexual psychotherapy, female therapist - female patient |
| 4012488 | Sexual psychotherapy, female therapist - male patient |
| 4132436 | Sexual psychotherapy, group |
| 4143316 | Sexual psychotherapy, group, all female |
| 4219683 | Sexual psychotherapy, group, all male |
| 4151904 | Sexual psychotherapy, group, male and female |
| 4278094 | Sexual psychotherapy, male therapist - female patient |
| 4249602 | Sexual psychotherapy, male therapist - male patient |
| 4234476 | Sexual surrogate therapy |
| 4179241 | Short-term psychodynamic therapy |
| 4234402 | Social psychotherapy |
| 4128406 | Specific task orientated psychotherapy |
| 4080044 | Stimulative psychotherapy |
| 4262582 | Structural family psychotherapy |
| 4263758 | Structural psychotherapy |
| 4126653 | Supportive expressive psychodynamic psychotherapy |
| 4311943 | Supportive verbal psychotherapy |
| 2007748 | Supportive verbal psychotherapy |
| 4225728 | Suppressive psychotherapy |
| 4080048 | Therapeutic psychology |
| 44808259 | Therapeutic role play |

**Table S1.3.** Concepts used to identify psychotherapy.

| Concept ID | Concept Name |
|---|---|
| 4111663 | Bilateral electroconvulsive therapy |
| 4030840 | Electroconvulsive therapy |
| 2108578 | Electroconvulsive therapy (ECT) provided (MDD) |
| 2213552 | Electroconvulsive therapy (includes necessary monitoring) |
| 4020981 | Electronarcosis |
| 4210144 | First treatment in a course of electroconvulsive therapy |
| 4336318 | Multiple electroconvulsive therapy |
| 4332436 | Multiple monitored electroconvulsive therapy |
| 2007728 | Other electroshock therapy |
| 44508134 | Other specified electroconvulsive therapy |
| 2108579 | Patient referral for electroconvulsive therapy (ECT) documented (MDD) |
| 2007727 | Subconvulsive electroshock therapy |
| 4004830 | Subconvulsive electroshock therapy |
| 4210145 | Subsequent treatment in a course of electroconvulsive therapy |

**Table S1.4.** Concepts used to identify electroconvulsive therapy.

## Document S2. Outcome definitions

This section describes the algorithms we use to identify occurrences of the 22 outcomes of interest. The algorithms are framed in the context of the OMOP Common Data Model[11] (CDM) version 5. The CDM uses a standardized terminology for encoding all information. For more information on the CDM and the standard vocabulary see http://ohdsi.org. The computer-executable version of these algorithms is part of the study R package: https://github.com/OHDSI/StudyProtocols/tree/master/LargeScalePopEst

### Acute liver injury

Note: This algorithm uses the set of codes identified by Udo et al. [12]

Initial Event Cohort

People having any of the following:

- a condition occurrence of acute liver injury[1]
  - for the first time in the person's history
  - visit occurrence is any of: Emergency Room Visit, Inpatient Visit

with continuous observation of at least 0 days prior and 0 days after event index date, and limit initial events to: **earliest event per person.**

For people matching the Primary Events, include:
Having all of the following criteria:

- exactly 0 occurrences of a condition occurrence of acute liver injury exclusion concepts[2]

  starting between 365 days Before and 60 days After event index date

Limit cohort of initial events to: **earliest event per person.**
Limit qualifying cohort to: **earliest event per person.**
No end date strategy selected. By default, the cohort end date will be the end of the observation period that contains the index event.

Appendix 1: Concept Set Definitions

1. acute liver injury

| Concept Id | Concept Name | Domain | Vocabulary | Excluded | Descendants | Mapped |
|---|---|---|---|---|---|---|
| 200763 | Chronic hepatitis | Condition | SNOMED | YES | YES | NO |
| 377604 | Hepatic coma | Condition | SNOMED | NO | YES | NO |

| Concept Id | Concept Name | Domain | Vocabulary | Excluded | Descendants | Mapped |
|---|---|---|---|---|---|---|
| 196029 | Hepatic coma due to viral hepatitis | Condition | SNOMED | YES | YES | NO |
| 4337543 | Hepatic necrosis | Condition | SNOMED | NO | YES | NO |
| 194087 | Hepatitis due to infection | Condition | SNOMED | YES | YES | NO |
| 196455 | Hepatorenal syndrome | Condition | SNOMED | NO | YES | NO |
| 194990 | Inflammatory disease of liver | Condition | SNOMED | NO | YES | NO |
| 4291005 | Viral hepatitis | Condition | SNOMED | YES | YES | NO |

2. acute liver injury exclusion concepts

| Concept Id | Concept Name | Domain | Vocabulary | Excluded | Descendants | Mapped |
|---|---|---|---|---|---|---|
| 192956 | Cholecystitis | Condition | SNOMED | NO | YES | NO |
| 200763 | Chronic hepatitis | Condition | SNOMED | NO | YES | NO |
| 4212540 | Chronic liver disease | Condition | SNOMED | NO | YES | NO |
| 197917 | Disorder of biliary tract | Condition | SNOMED | NO | YES | NO |
| 192353 | Disorder of gallbladder | Condition | SNOMED | NO | YES | NO |
| 192963 | Disorder of pancreas | Condition | SNOMED | NO | YES | NO |
| 196456 | Gallstone | Condition | SNOMED | NO | YES | NO |
| 4130518 | Neoplasm of liver | Condition | SNOMED | NO | YES | NO |
| 4291005 | Viral hepatitis | Condition | SNOMED | NO | YES | NO |

## Acute myocardial infarction

Initial Event Cohort

People having any of the following:

- a condition occurrence of Acute MI[1]
  - for the first time in the person's history
  - condition type is any of: Inpatient detail - primary, Inpatient header - primary, Primary Condition, Inpatient detail - 1st position, Inpatient header - 1st position
  - visit occurrence is any of: Emergency Room Visit, Inpatient Visit

with continuous observation of at least 0 days prior and 0 days after event index date, and limit initial events to: **earliest event per person.**

Limit qualifying cohort to: **earliest event per person.**
No end date strategy selected. By default, the cohort end date will be the end of the observation period that contains the index event.

Appendix 1: Concept Set Definitions

1. Acute MI

| Concept Id | Concept Name | Domain | Vocabulary | Excluded | Descendants | Mapped |
|---|---|---|---|---|---|---|
| 4329847 | Myocardial infarction | Condition | SNOMED | NO | YES | NO |
| 314666 | Old myocardial infarction | Condition | SNOMED | YES | YES | NO |

## Alopecia

Initial Event Cohort

People having any of the following:

- a condition occurrence of Alopecia[1]
  - for the first time in the person's history

with continuous observation of at least 0 days prior and 0 days after event index date, and limit initial events to: **earliest event per person.**

Limit qualifying cohort to: **earliest event per person.**
No end date strategy selected. By default, the cohort end date will be the end of the observation period that contains the index event.

Appendix 1: Concept Set Definitions

1. Alopecia

| Concept Id | Concept Name | Domain | Vocabulary | Excluded | Descendants | Mapped |
|---|---|---|---|---|---|---|
| 133280 | Alopecia | Condition | SNOMED | NO | YES | NO |
| 133959 | Syphilitic alopecia | Condition | SNOMED | YES | YES | NO |

## Constipation

Note: This algorithm requires the occurrence of 2 or more diagnoses, as recommended by Mody et al. [13]

Initial Event Cohort

People having any of the following:

- a condition occurrence of Constipation[1]
  - for the first time in the person's history

with continuous observation of at least 0 days prior and 0 days after event index date, and limit initial events to: **earliest event per person.**

Limit qualifying cohort to: **earliest event per person.**
No end date strategy selected. By default, the cohort end date will be the end of the observation period that contains the index event.

Appendix 1: Concept Set Definitions

1. Constipation

| Concept Id | Concept Name | Domain | Vocabulary | Excluded | Descendants | Mapped |
|---|---|---|---|---|---|---|
| 75860 | Constipation | Condition | SNOMED | NO | YES | NO |

## Decreased libido

Initial Event Cohort

People having any of the following:

- a condition occurrence of Decreased libido[1]
  - for the first time in the person's history

with continuous observation of at least 0 days prior and 0 days after event index date, and limit initial events to: **earliest event per person.**

Limit qualifying cohort to: **earliest event per person.**

No end date strategy selected. By default, the cohort end date will be the end of the observation period that contains the index event.

Appendix 1: Concept Set Definitions

1. Decreased libido

| Concept Id | Concept Name | Domain | Vocabulary | Excluded | Descendants | Mapped |
|---|---|---|---|---|---|---|
| 436246 | Reduced libido | Condition | SNOMED | NO | YES | NO |

## Delirium

Note: This algorithm relies on diagnosis codes associated with hospitalization. This approach may lead to underreporting, as described by McCoy et al. [14]

Initial Event Cohort

People having any of the following:

- a condition occurrence of Delirium[1]
  - for the first time in the person's history
  - visit occurrence is any of: Emergency Room Visit, Inpatient Visit

with continuous observation of at least 0 days prior and 0 days after event index date, and limit initial events to: **earliest event per person.**

Limit qualifying cohort to: **earliest event per person.**

No end date strategy selected. By default, the cohort end date will be the end of the observation period that contains the index event.

Appendix 1: Concept Set Definitions

1. Delirium

| Concept Id | Concept Name | Domain | Vocabulary | Excluded | Descendants | Mapped |
|---|---|---|---|---|---|---|
| 377830 | Alcohol withdrawal delirium | Condition | SNOMED | YES | YES | NO |
| 373995 | Delirium | Condition | SNOMED | NO | YES | NO |

## Diarrhea

Note: This algorithm follows Broder et al. [15]

Initial Event Cohort

People having any of the following:

- a condition occurrence of Diarrhea[1]
  - for the first time in the person's history

with continuous observation of at least 0 days prior and 0 days after event index date, and limit initial events to: **earliest event per person.**

Limit qualifying cohort to: **earliest event per person.**
No end date strategy selected. By default, the cohort end date will be the end of the observation period that contains the index event.

Appendix 1: Concept Set Definitions

1. Diarrhea

| Concept Id | Concept Name | Domain | Vocabulary | Excluded | Descendants | Mapped |
|---|---|---|---|---|---|---|
| 196523 | Diarrhea | Condition | SNOMED | NO | YES | NO |
| 80141 | Functional diarrhea | Condition | SNOMED | NO | YES | NO |

## Fracture

Note: This algorithm follows Lanteigne et al. [16]

Initial Event Cohort

People having any of the following:

- a condition occurrence of Fracture[1]
  - for the first time in the person's history

with continuous observation of at least 0 days prior and 0 days after event index date, and limit initial events to: **earliest event per person.**

Limit qualifying cohort to: **earliest event per person.**
No end date strategy selected. By default, the cohort end date will be the end of the observation period that contains the index event.

Appendix 1: Concept Set Definitions

1. Fracture

| Concept Id | Concept Name | Domain | Vocabulary | Excluded | Descendants | Mapped |
|---|---|---|---|---|---|---|
| 435093 | Closed fracture of femur | Condition | SNOMED | NO | YES | NO |
| 441974 | Closed fracture of forearm | Condition | SNOMED | NO | YES | NO |
| 4230399 | Closed fracture of hip | Condition | SNOMED | NO | YES | NO |
| 441422 | Closed fracture of humerus | Condition | SNOMED | NO | YES | NO |
| 439166 | Closed fracture of radius | Condition | SNOMED | NO | YES | NO |
| 4278672 | Fracture of forearm | Condition | SNOMED | NO | YES | NO |
| 442619 | Fracture of humerus | Condition | SNOMED | NO | YES | NO |
| 433856 | Fracture of neck of femur | Condition | SNOMED | NO | YES | NO |
| 4131595 | Fracture of radius | Condition | SNOMED | NO | YES | NO |
| 73571 | Pathological fracture | Condition | SNOMED | NO | YES | NO |

## Gastrointestinal hemhorrage

Initial Event Cohort

People having any of the following:

- a condition occurrence of Gastrointestinal hemorrhage[1]
  - for the first time in the person's history
  - condition type is any of: Inpatient detail - primary, Inpatient header - primary, Primary Condition, Inpatient detail - 1st position, Inpatient header - 1st position
  - visit occurrence is any of: Emergency Room Visit, Inpatient Visit

with continuous observation of at least 0 days prior and 0 days after event index date, and limit initial events to: **earliest event per person.**

Limit qualifying cohort to: **earliest event per person.**
No end date strategy selected. By default, the cohort end date will be the end of the observation period that contains the index event.

Appendix 1: Concept Set Definitions

1. Gastrointestinal hemorrhage

| Concept Id | Concept Name | Domain | Vocabulary | Excluded | Descendants | Mapped |
|---|---|---|---|---|---|---|
| 4280942 | Acute gastrojejunal ulcer with perforation | Condition | SNOMED | NO | YES | NO |
| 28779 | Bleeding esophageal varices | Condition | SNOMED | NO | YES | NO |
| 198798 | Dieulafoy's vascular malformation | Condition | SNOMED | NO | YES | NO |
| 4112183 | Esophageal varices with bleeding, associated with another disorder | Condition | SNOMED | NO | YES | NO |

| Concept Id | Concept Name | Domain | Vocabulary | Excluded | Descendants | Mapped |
|---|---|---|---|---|---|---|
| 194382 | External hemorrhoids | Condition | SNOMED | NO | NO | NO |
| 192671 | Gastrointestinal hemorrhage | Condition | SNOMED | NO | YES | NO |
| 196436 | Internal hemorrhoids | Condition | SNOMED | NO | NO | NO |
| 4338225 | Peptic ulcer with perforation | Condition | SNOMED | NO | YES | NO |
| 194158 | Perinatal gastrointestinal hemorrhage | Condition | SNOMED | YES | YES | NO |

## Hyperprolactinemia

Initial Event Cohort

People having any of the following:

- a condition occurrence of Hyperprolactinemia[1]
    - for the first time in the person's history

with continuous observation of at least 0 days prior and 0 days after event index date, and limit initial events to: **earliest event per person.**

Limit qualifying cohort to: **earliest event per person.**
No end date strategy selected. By default, the cohort end date will be the end of the observation period that contains the index event.

Appendix 1: Concept Set Definitions

1. Hyperprolactinemia

| Concept Id | Concept Name | Domain | Vocabulary | Excluded | Descendants | Mapped |
|---|---|---|---|---|---|---|
| 4030186 | Hyperprolactinemia | Condition | SNOMED | NO | YES | NO |

## Hyponatremia

Note: The algorithm here relies on the recording of diagnoses codes, and might not have high sensitivity as remarked by Shea et al.

Initial Event Cohort

People having any of the following:

- a condition occurrence of Hyponatremia[1]
    - for the first time in the person's history
- a measurement of Serum sodium[2]
    - for the first time in the person's history
    - with value as number $< 136$
    - unit is any of: millimole per liter

with continuous observation of at least 0 days prior and 0 days after event index date, and limit initial events to: **earliest event per person.**

Limit qualifying cohort to: **earliest event per person.**
No end date strategy selected. By default, the cohort end date will be the end of the observation period that contains the index event.

Appendix 1: Concept Set Definitions

1. Hyponatremia

| Concept Id | Concept Name | Domain | Vocabulary | Excluded | Descendants | Mapped |
|---|---|---|---|---|---|---|
| 435515 | Hypo-osmolality and or hyponatremia | Condition | SNOMED | NO | YES | NO |

2. Serum sodium

| Concept Id | Concept Name | Domain | Vocabulary | Excluded | Descendants | Mapped |
|---|---|---|---|---|---|---|
| 3032987 | Sodium [Moles/volume] corrected for glucose in Serum or Plasma | Measurement | LOINC | NO | YES | NO |
| 46235784 | Sodium [Moles/volume] in Serum, Plasma or Blood | Measurement | LOINC | NO | YES | NO |
| 3019550 | Sodium serum/plasma | Measurement | LOINC | NO | YES | NO |

## Hypotension

Note: This algorithm follows Wernli et al. [17]

Initial Event Cohort
People having any of the following:

- a condition occurrence of Hypotension[1]
    - for the first time in the person's history

with continuous observation of at least 0 days prior and 0 days after event index date, and limit initial events to: **earliest event per person.**

Limit qualifying cohort to: **earliest event per person.**
No end date strategy selected. By default, the cohort end date will be the end of the observation period that contains the index event.

Appendix 1: Concept Set Definitions

1. Hypotension

| Concept Id | Concept Name | Domain | Vocabulary | Excluded | Descendants | Mapped |
|---|---|---|---|---|---|---|
| 4120275 | Drug-induced hypotension | Condition | SNOMED | NO | YES | NO |
| 317002 | Low blood pressure | Condition | SNOMED | NO | YES | NO |

| Concept Id | Concept Name | Domain | Vocabulary | Excluded | Descendants | Mapped |
|---|---|---|---|---|---|---|
| 314432 | Maternal hypotension syndrome | Condition | SNOMED | YES | YES | NO |
| 319041 | Orthostatic hypotension | Condition | SNOMED | NO | YES | NO |

## Hypothyroidism

Note: This algorithm requires the occurrences of 2 more diagnose codes, as recommended by Lu et al. [18]

Initial Event Cohort

People having any of the following:

- a condition occurrence of Hypothyroidism[1]

with continuous observation of at least 0 days prior and 0 days after event index date, and limit initial events to: **all events per person.**

For people matching the Primary Events, include:
Having all of the following criteria:

- at least 2 occurrences of a condition occurrence of Hypothyroidism[1]

  starting between 0 days Before and 90 days After event index date

Limit cohort of initial events to: **earliest event per person.**
Limit qualifying cohort to: **earliest event per person.**
No end date strategy selected. By default, the cohort end date will be the end of the observation period that contains the index event.

Appendix 1: Concept Set Definitions

1. Hypothyroidism

| Concept Id | Concept Name | Domain | Vocabulary | Excluded | Descendants | Mapped |
|---|---|---|---|---|---|---|
| 140673 | Hypothyroidism | Condition | SNOMED | NO | YES | NO |

## Insomnia

Initial Event Cohort

People having any of the following:

- a condition occurrence of Insomnia[1]
  - for the first time in the person's history

with continuous observation of at least 0 days prior and 0 days after event index date, and limit initial events to: **earliest event per person.**

Limit qualifying cohort to: **earliest event per person.**
No end date strategy selected. By default, the cohort end date will be the end of the observation period that contains the index event.

Appendix 1: Concept Set Definitions

1. Insomnia

| Concept Id | Concept Name | Domain | Vocabulary | Excluded | Descendants | Mapped |
|---|---|---|---|---|---|---|
| 439708 | Disorders of initiating and maintaining sleep | Condition | SNOMED | NO | YES | NO |
| 436962 | Insomnia | Condition | SNOMED | NO | YES | NO |
| 4305303 | Sleep deprivation | Condition | SNOMED | NO | YES | NO |

## Nausea

Initial Event Cohort
People having any of the following:

- a condition occurrence of Nausea[1]
  - for the first time in the person's history

with continuous observation of at least 0 days prior and 0 days after event index date, and limit initial events to: **earliest event per person.**

Limit qualifying cohort to: **earliest event per person.**
No end date strategy selected. By default, the cohort end date will be the end of the observation period that contains the index event.

Appendix 1: Concept Set Definitions

1. Nausea

| Concept Id | Concept Name | Domain | Vocabulary | Excluded | Descendants | Mapped |
|---|---|---|---|---|---|---|
| 30284 | Motion sickness | Condition | SNOMED | YES | YES | NO |
| 31967 | Nausea | Condition | SNOMED | NO | YES | NO |

## Open-angle glaucoma

Note: This algorithm follows Stein et al. [19]

Initial Event Cohort
People having any of the following:

- a condition occurrence of Open-angle glaucoma[1]
  - for the first time in the person's history

with continuous observation of at least 365 days prior and 0 days after event index date, and limit initial events to: **earliest event per person.**

For people matching the Primary Events, include:
Having all of the following criteria:

- at least 1 occurrences of a condition occurrence of Open-angle glaucoma[1]
  - provider specialty is any of: Ophthalmology, Optometry, Optician

  starting between 1 days After and 365 days After event index date

Limit cohort of initial events to: **earliest event per person.**
Limit qualifying cohort to: **all events per person.**
No end date strategy selected. By default, the cohort end date will be the end of the observation period that contains the index event.

Appendix 1: Concept Set Definitions

1. Open-angle glaucoma

| Concept Id | Concept Name | Domain | Vocabulary | Excluded | Descendants | Mapped |
|---|---|---|---|---|---|---|
| 432908 | Glaucomatocyclitic crisis | Condition | SNOMED | YES | YES | NO |
| 441561 | Low tension glaucoma | Condition | SNOMED | NO | YES | NO |
| 4216823 | Open angle with borderline findings | Condition | SNOMED | YES | YES | NO |
| 441284 | Open-angle glaucoma | Condition | SNOMED | NO | YES | NO |
| 4072218 | Secondary open-angle glaucoma | Condition | SNOMED | YES | YES | NO |

## Seizure

Note: This algorithm requires either inpatient or emergency room visits as recommended by Wu et al. [20]

Initial Event Cohort
People having any of the following:

- a condition occurrence of Seizure and seizure disorder[1]
  - for the first time in the person's history
  - visit occurrence is any of: Emergency Room Visit, Inpatient Visit

with continuous observation of at least 0 days prior and 0 days after event index date, and limit initial events to: **earliest event per person.**

Limit qualifying cohort to: **earliest event per person.**
No end date strategy selected. By default, the cohort end date will be the end of the observation period that contains the index event.

Appendix 1: Concept Set Definitions

1. Seizure and seizure disorder

| Concept Id | Concept Name | Domain | Vocabulary | Excluded | Descendants | Mapped |
|---|---|---|---|---|---|---|
| 380533 | Convulsions in the newborn | Condition | SNOMED | YES | YES | NO |
| 45757050 | Epilepsy in mother complicating pregnancy | Condition | SNOMED | YES | YES | NO |
| 377091 | Seizure | Condition | SNOMED | NO | YES | NO |
| 4029498 | Seizure disorder | Condition | SNOMED | NO | YES | NO |

## Stroke

Initial Event Cohort

People having any of the following:

- a condition occurrence of Ischemic stroke[1]
  - for the first time in the person's history
  - visit occurrence is any of: Inpatient Visit

with continuous observation of at least 0 days prior and 0 days after event index date, and limit initial events to: **earliest event per person.**

Limit qualifying cohort to: **earliest event per person.**
No end date strategy selected. By default, the cohort end date will be the end of the observation period that contains the index event.

Appendix 1: Concept Set Definitions

1. Ischemic stroke

| Concept Id | Concept Name | Domain | Vocabulary | Excluded | Descendants | Mapped |
|---|---|---|---|---|---|---|
| 374060 | Acute ill-defined cerebrovascular disease | Condition | SNOMED | NO | YES | NO |
| 4108356 | Cerebral infarction due to embolism of cerebral arteries | Condition | SNOMED | NO | YES | NO |
| 4110192 | Cerebral infarction due to thrombosis of cerebral arteries | Condition | SNOMED | NO | YES | NO |
| 4043731 | Infarction - precerebral | Condition | SNOMED | NO | YES | NO |

## Suicide and suicidal ideation

Note: This algorithm is based on the review by Callagan et al. [21]

Initial Event Cohort

People having any of the following:

- a condition occurrence of Suicide and suicidal ideation[1]
    - for the first time in the person's history
- an observation of Suicide and suicidal ideation[1]
    - for the first time in the person's history

with continuous observation of at least 0 days prior and 0 days after event index date, and limit initial events to: **earliest event per person.**

Limit qualifying cohort to: **earliest event per person.**
No end date strategy selected. By default, the cohort end date will be the end of the observation period that contains the index event.

Appendix 1: Concept Set Definitions

1. Suicide and suicidal ideation

| Concept Id | Concept Name | Domain | Vocabulary | Excluded | Descendants | Mapped |
|---|---|---|---|---|---|---|
| 439235 | Self inflicted injury | Condition | SNOMED | NO | YES | NO |
| 4181216 | Self-administered poisoning | Condition | SNOMED | NO | YES | NO |
| 444362 | Suicidal deliberate poisoning | Condition | SNOMED | NO | YES | NO |
| 4273391 | Suicidal thoughts | Condition | SNOMED | NO | YES | NO |
| 440925 | Suicide | Observation | SNOMED | NO | YES | NO |

## Tinnitus

Note: This algorithm follows Lee et al. [22]

Initial Event Cohort
People having any of the following:

- a condition occurrence of Tinnitus[1]
    - for the first time in the person's history

with continuous observation of at least 0 days prior and 0 days after event index date, and limit initial events to: **earliest event per person.**

Limit qualifying cohort to: **earliest event per person.**
No end date strategy selected. By default, the cohort end date will be the end of the observation period that contains the index event.

Appendix 1: Concept Set Definitions

1. Tinnitus

| Concept Id | Concept Name | Domain | Vocabulary | Excluded | Descendants | Mapped |
|---|---|---|---|---|---|---|
| 377575 | Tinnitus | Condition | SNOMED | NO | YES | NO |

## Ventricular arrhythmia and sudden cardiac death

Note: This algorithm follows the definition used by Leonard et al. [23]

Initial Event Cohort

People having any of the following:

- a condition occurrence of Ventricular arrhythmia and sudden cardiac death[1]
    - for the first time in the person's history
    - condition type is any of: Inpatient detail - primary, Inpatient header - primary, Primary Condition, Carrier claim detail - 1st position, Carrier claim header - 1st position, Inpatient detail - 1st position, Inpatient header - 1st position, Outpatient detail - 1st position, Outpatient header - 1st position
    - visit occurrence is any of: Emergency Room Visit, Inpatient Visit

with continuous observation of at least 0 days prior and 0 days after event index date, and limit initial events to: **earliest event per person.**

Limit qualifying cohort to: **earliest event per person.**
No end date strategy selected. By default, the cohort end date will be the end of the observation period that contains the index event.

Appendix 1: Concept Set Definitions

1. Ventricular arrhythmia and sudden cardiac death

| Concept Id | Concept Name | Domain | Vocabulary | Excluded | Descendants | Mapped |
|---|---|---|---|---|---|---|
| 321042 | Cardiac arrest | Condition | SNOMED | NO | YES | NO |
| 442289 | Death in less than 24 hours from onset of symptoms | Observation | SNOMED | NO | YES | NO |
| 441139 | Instantaneous death | Observation | SNOMED | NO | YES | NO |
| 4132309 | Sudden death | Observation | SNOMED | NO | YES | NO |
| 4185572 | Ventricular arrhythmia | Condition | SNOMED | NO | YES | NO |
| 437894 | Ventricular fibrillation | Condition | SNOMED | NO | YES | NO |
| 4103295 | Ventricular tachycardia | Condition | SNOMED | NO | YES | NO |

## Vertigo

Initial Event Cohort

People having any of the following:

- a condition occurrence of Vertigo[1]
    - for the first time in the person's history

with continuous observation of at least 0 days prior and 0 days after event index date, and limit initial events to: **earliest event per person.**

Limit qualifying cohort to: **earliest event per person.**
No end date strategy selected. By default, the cohort end date will be the end of the observation period that contains the index event.

Appendix 1: Concept Set Definitions

1. Vertigo

| Concept Id | Concept Name | Domain | Vocabulary | Excluded | Descendants | Mapped |
|---|---|---|---|---|---|---|
| 78162 | Peripheral vertigo | Condition | SNOMED | NO | YES | NO |
| 439383 | Vertigo | Condition | SNOMED | NO | YES | NO |
| 381035 | Vertigo of central origin | Condition | SNOMED | NO | YES | NO |

## Document S3. Consistency with gold standard from randomized controlled trials

Sertraline increases risk of diarrhea relative to comparators: Across all four databases, we consistently observe sertraline having an increased risk of diarrhea relative to most comparator treatments, with the magnitude of effect estimates ranging from 20% to 100% increased risk. In MDCR, we produce estimates for 11 comparisons with sertraline, 10 of which yield calibrated 95% confidence intervals greater than 1: nortriptyline HR=2.10 (95% CI: 1.51-2.91); psychotherapy HR=1.96 (95% CI: 1.22-3.41); fluoxetine HR=1.71 (95% CI: 1.34-2.20); sertraline HR=1.68 (95% CI: 1.44-1.97); venlafaxine HR=1.58 (95% CI: 1.35-1.86); amitriptyline HR=1.56 (95% CI: 1.10-2.37); duloxetine HR=1.33 (95% CI: 1.09-1.68); trazodone HR=1.33 (95% CI: 1.12-1.59); citalopram HR=1.31 (95% CI: 1.15-1.49); escitalopram HR=1.22 (95% CI: 1.07-1.40); and mirtazapine HR=1.16 (95% CI: 0.98-1.38). All comparisons in MDCR except psychotherapy pass all empirical diagnostics, with sufficient sample near clinical equipoise, adequate covariate balance, and empirical calibration demonstrating nominal operating characteristics. For the comparison with psychotherapy, inadequate covariate balance remains after propensity score adjustment, suggesting the potential for residual bias.

Venlafaxine increases risk of nausea relative to SSRIs: Amongst the two privately-insured populations (CCAE and Optum), we find consistent evidence of a small increased risk of nausea between new users of venlafaxine and the most prevalent SSRIs – sertraline, escitalopram, and citalopram. When comparing the risk of nausea between new users of venlafaxine and sertraline in the CCAE database, the estimated calibrated HR is 1.07 (95% CI: 0.91-1.26), and Optum returns a similar effect estimate [HR=1.10 (95% CI: 1.00-1.22)]. When comparing the risk of nausea between new users of venlafaxine and escitalopram, the estimated calibrated HR in CCAE is HR=1.12 (95% CI: 1.01-1.25) and in Optum is HR=1.12 (95% CI: 0.97-1.30). When comparing the risk of nausea between new users of venlafaxine and citalopram, the estimated calibrated HR in CCAE is HR=1.08 (95% CI: 1.00-1.18) and in Optum is HR=1.05 (95% CI: 0.95-1.18). Across all databases, comparisons between venlafaxine and each SSRI demonstrate sufficient clinical equipoise, adequate covariate balance, and empirical calibration restores nominal operating characteristics. Compared with the clinical trial results, the observational studies have a lower incidence of nausea, and are directionally consistent but had a small magnitude of effect estimate.

No difference in nausea between duloxetine and paroxetine or fluoxetine: All four databases produce estimates comparing the duloxetine with fluoxetine, all of which consistently suggested no association. In CCAE, when comparing the risk of nausea between new users of duloxetine and fluoxetine, the estimated calibrated HR is 0.93 (95% CI: 0.79-1.12). Relative to paroxetine, the estimated calibrated HR is 1.01 (95% CI: 0.85-1.19). These observational studies allow the use of large samples to bound the magnitude of any potential effect to increase confidence in any conclusion around non-inferiority.

Paroxetine higher rate of sexual dysfunction than fluoxetine and sertraline: Among clinical trials of SSRIs, paroxetine is observed to have non-significant higher rates of sexual dysfunction than fluoxetine and sertraline. Across our observational databases, paroxetine is the least commonly used SSRI. In our largest dataset (CCAE), there are <8000 new users used in each analysis. While the review compared treatments for the broadly defined ‘sexual dysfunction’ outcome, we estimate effects for a more narrowly defined diagnosis of ‘decreased libido’, and the incidence of events is lower in observational data than reported in the trials. When comparing the risk of decreased libido between new users of paroxetine and fluoxetine, the estimated calibrated HR is 1.40 (95% CI: 0.84-2.34), and in comparison with sertraline, the estimated calibrated HR is 1.39 (95% CI: 0.86-2.26).

## Figure S1. Estimates for all control hypotheses before and after calibration

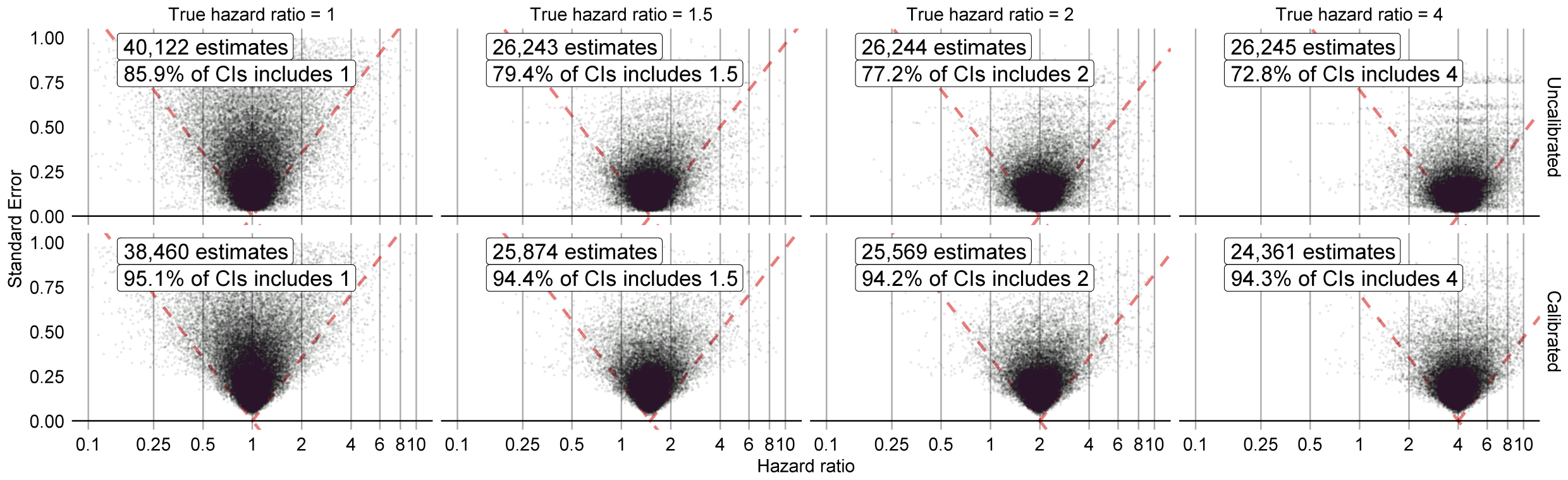

**Figure S1**. Hazard ratios and corresponding standard errors estimated through our systematic evidence generation process to four observational databases for our negative and positive controls before (top) and after (bottom) calibration. The estimates are stratified by the true hazard ratio. Note that due to limitations in sample size not all negative controls could be used to synthesize positive controls, and a small fraction of estimates could therefore not be calibrated.

## Figure S2. Leave-one-out cross-validation of calibration

To validate our confidence interval calibration procedure we use a leave-one-out cross-validation. For each negative control and the positive controls derived from that negative control, we fit systematic error models using all other controls, and compute confidence intervals for the left-out controls with a wide range of widths. We subsequently check how often the confidence intervals contained the true hazard ratio. In each fold of the cross-validation, error models are computed separately for each combination of target, comparator, and database. Figure S3 shows the coverage as a function of width of the confidence interval, stratified by the true hazard ratio.

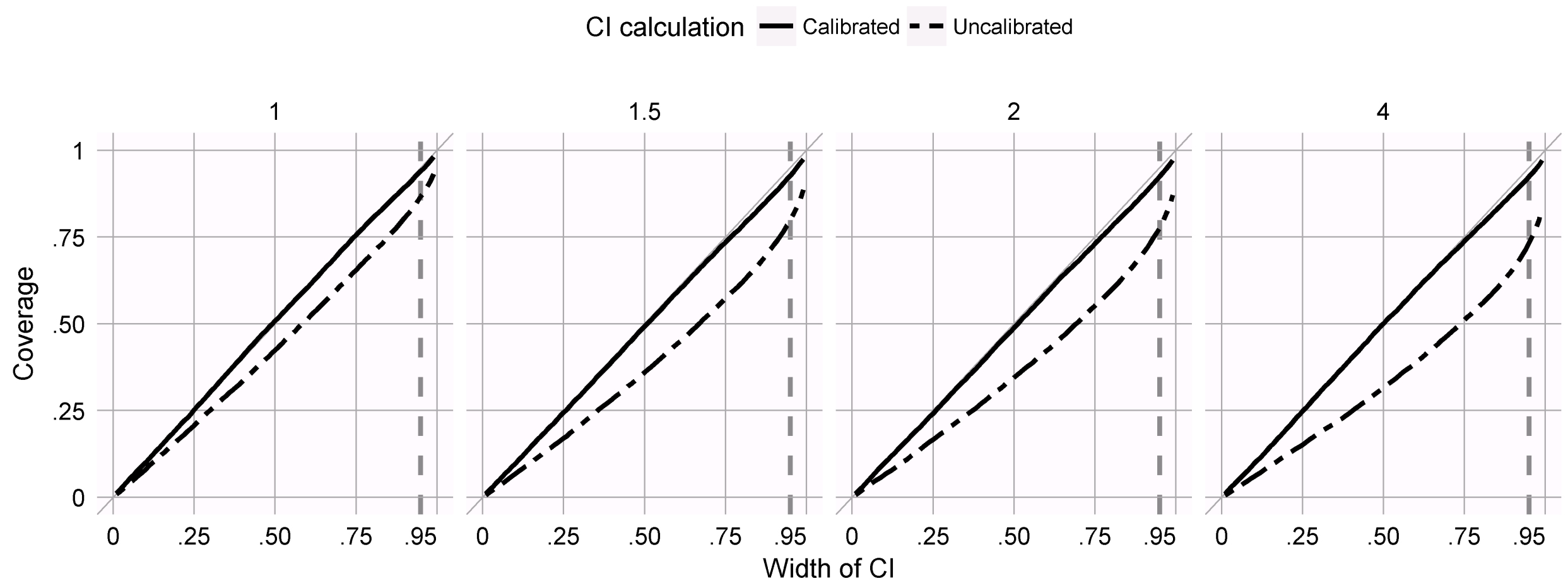

**Figure S2.** Coverage of confidence intervals, per width of the confidence interval, and stratified by true hazard ratio.

# Figure S3. Comparison of results from RCTs and our observational study

| Source | Comparator | Effect size (95% CI) |
|---|---|---|
| Coleman 1999 | Bupropion | 1.67 (0.81-3.46) |
| Croft 1999 | Bupropion | 4.93 (2.16-11.26) |
| Kavoussi 1997 | Bupropion | 8.43 (2.86-24.85) |
| CCAE | Bupropion | 1.61 (1.29-2.04) |
| MDCD | Bupropion | 1.30 (1.00-1.68) |
| MDCR | Bupropion | 1.68 (1.44-1.96) |
| Optum | Bupropion | 1.55 (1.28-1.90) |
| Ekselius 1997 | Citalopram | 1.54 (0.92-2.56) |
| CCAE | Citalopram | 1.28 (1.21-1.36) |
| MDCD | Citalopram | 1.06 (0.91-1.26) |
| MDCR | Citalopram | 1.31 (1.15-1.49) |
| Optum | Citalopram | 1.30 (1.18-1.44) |
| Alexopoulos 2004 | Escitalopram | 2.01 (0.90-4.50) |
| Ventura 2007 | Escitalopram | 2.18 (1.05-4.53) |
| CCAE | Escitalopram | 1.22 (1.05-1.43) |
| MDCD | Escitalopram | 1.25 (1.06-1.46) |
| MDCR | Escitalopram | 1.22 (1.06-1.41) |
| Optum | Escitalopram | 1.23 (1.11-1.38) |
| Bennie 1995 | Fluoxetine | 1.44 (0.45-4.65) |
| Fava 2002 | Fluoxetine | 1.55 (0.77-3.12) |
| Newhouse 2000 | Fluoxetine | 1.50 (0.78-2.90) |
| Sechter 1999 | Fluoxetine | 1.54 (0.25-9.38) |
| CCAE | Fluoxetine | 1.19 (1.09-1.30) |
| MDCD | Fluoxetine | 1.02 (0.89-1.18) |
| MDCR | Fluoxetine | 1.71 (1.34-2.20) |
| Optum | Fluoxetine | 1.15 (1.06-1.25) |
| Behnke 2003 | Mirtazapine | 2.51 (1.01-6.26) |
| Thase 2000 | Mirtazapine | 2.93 (1.34-6.40) |
| CCAE | Mirtazapine | 1.20 (0.92-1.65) |
| MDCD | Mirtazapine | 1.02 (0.84-1.25) |
| MDCR | Mirtazapine | 1.16 (0.98-1.38) |
| Optum | Mirtazapine | 1.23 (1.00-1.56) |
| Munizza 2006 | Trazodone | 1.58 (0.25-9.80) |
| CCAE | Trazodone | 1.23 (1.05-1.45) |
| MDCD | Trazodone | 1.01 (0.86-1.18) |
| MDCR | Trazodone | 1.33 (1.12-1.59) |
| Optum | Trazodone | 1.14 (1.01-1.29) |
| Mehtonen 2000 | Venlafaxine | 1.85 (0.64-5.40) |
| Shelton 2006 | Venlafaxine | 1.22 (0.60-2.51) |
| CCAE | Venlafaxine | 1.42 (1.20-1.70) |
| MDCD | Venlafaxine | 1.16 (0.95-1.42) |
| MDCR | Venlafaxine | 1.58 (1.35-1.85) |
| Optum | Venlafaxine | 1.38 (1.23-1.55) |

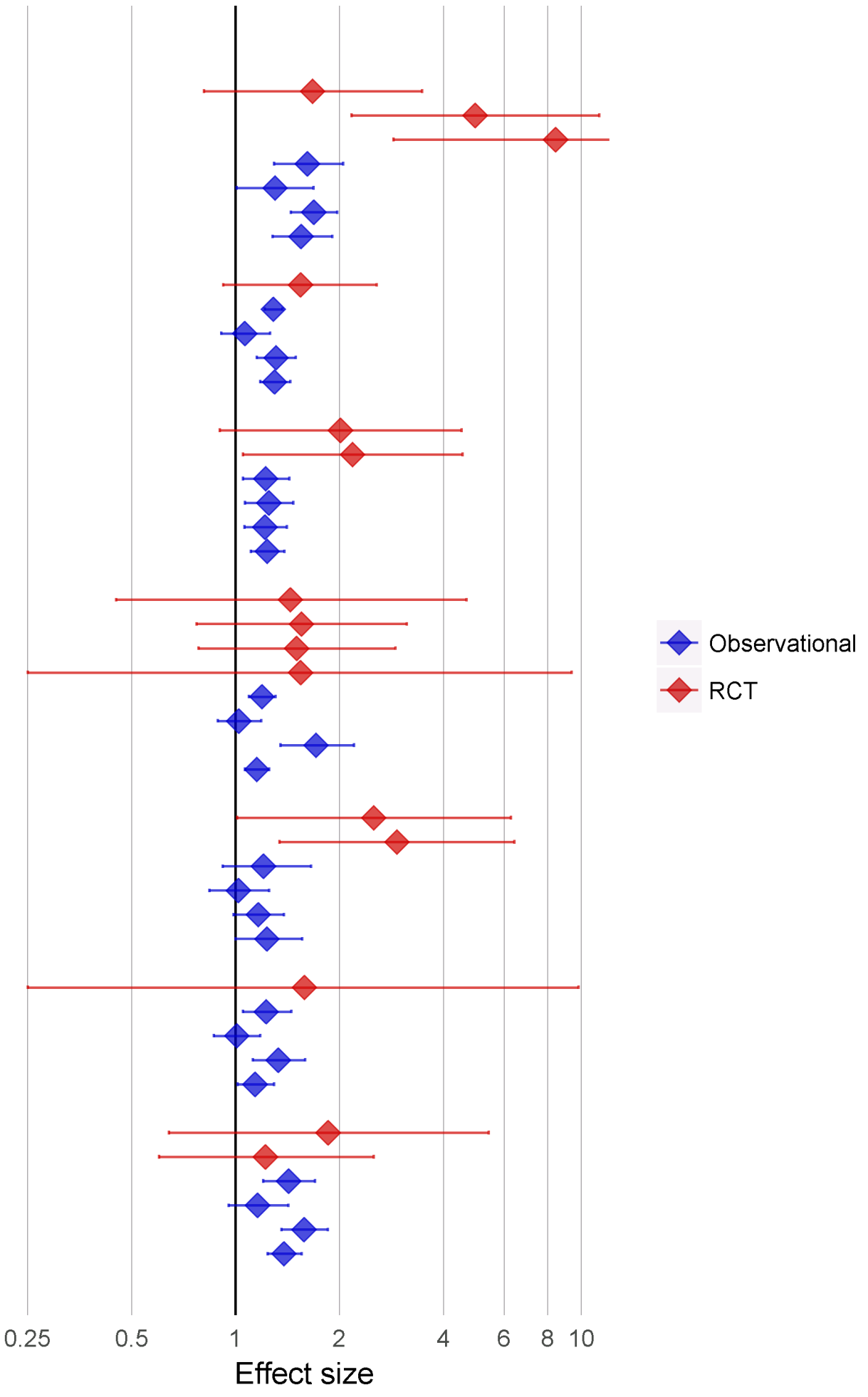

**Figure S3.** Effect size estimates on Sertraline compared to various other drugs for the outcome of diarrhea. Estimates from RCTs are those reported in Cipriani et al (2010). Observational estimates are the hazard ratios and 95% calibrated confidence intervals generated in our study.